\documentclass[prb,twocolumn,showpacs,superscriptaddress]{revtex4-1}

\usepackage{graphicx}
\usepackage{amsmath,amssymb,bm}

\usepackage{color}

\begin{document}


\title{
Exact interacting Green's function for the Anderson impurity 
at high bias voltages
}


\author{Akira Oguri}
\affiliation{
Department of Physics, Osaka City University, Sumiyoshi-ku, 
Osaka 558-8585, Japan
}

\author{Rui Sakano}
\affiliation{
Institute for Solid State Physics, University of Tokyo, Kashiwa, Chiba 277-8581, Japan
}

\date{\today}


\begin{abstract}
We describe some exact high-energy properties of 
a single Anderson impurity connected to two noninteracting leads 
in a nonequilibrium steady state.
In the limit of high bias voltages, 
and also in the high-temperature limit at thermal equilibrium, 
 the model can be mapped on to 
an effective non-Hermitian Hamiltonian consisting of two sites, 
which correspond  to the original impurity and its image 
that is defined  in a doubled Hilbert space 
referred to as Liouville-Fock space. 
For this, we provide a heuristic derivation  
using a path-integral representation 
of the Keldysh contour and the thermal field theory,  
in which the time evolution along the backward contour  
is replicated by extra degrees of freedom corresponding to the image.
We find that the effective Hamiltonian can also be expressed  
in terms of charges and currents. 
From this, it can be deduced that 
the dynamic susceptibilities for the charges 
and the current fluctuations become independent of the Coulomb repulsion $U$ 
in the high bias limit.
Furthermore, the equation of motions for the Green's function 
and 
two other higher-order correlation functions 
constitute a closed system.
The exact solution obtained from the three coupled equations extends  
the atomic-limit solution such that   
the self-energy correctly captures the imaginary part 
caused by the relaxation processes at high energies.  
The spectral weights of the upper and lower Hubbard levels 
depend sensitively on the asymmetry in the tunneling couplings 
to the left and right leads.

\end{abstract}


\maketitle

\section{Introduction}
\label{sec:introduction}

The impurity Anderson model is one of the most fundamental 
models for strongly correlated electrons 
in dilute magnetic alloys,\cite{Anderson}
quantum dots,\cite{HDW,MW} and also bulk systems 
in conjunction with the dynamical mean-field theory.\cite{DMFT} 
The model captures essential physics of 
 quantum  impurities for the whole energy regions: 
namely from the formation of a local magnetic moment 
and the Coulomb blockade occurring at high energies 
to the low-energy Kondo screening of the local moment.\cite{Hewson_book}

The Kondo effect has  been studied intensively   
for quantum dots, especially in a nonequilibrium  steady state  
driven by an applied bias voltage.\cite{HDW,MW}
For instance, the universal Kondo behavior 
of the steady current\cite{KNG,ao2001,HBA} and the 
shot noise\cite{GogolinKomnik,Sela2006,Golub,Mora2009,Fujii2010,Sakano} 
have been shown to be determined by 
the local-Fermi-liquid parameters for  
quasiparticles at low energies, 
where both the bias voltage $eV$ and 
 temperature $T$ are much smaller than the Kondo energy scale $T_K$.
Low-temperature experiments  have also been carried out 
for the steady current\cite{Grobis,ScottNatelson} 
and the shot noise\cite{Heiblum,Kobayashi} 
to examine the universal behavior.

In contrast to the low-energy regions, 
the properties in the higher energy regions beyond 
the Fermi-liquid regime have still not been fully understood. 
In order to explore the intermediate energy regions, 
numerical approaches such as 
the Wilson numerical renormalization group,\cite{Anders} 
the density matrix renormalization group,\cite{Kirino} 
the continuous-time quantum Monte Carlo methods,
\cite{Werner,MuhlbacherUrbanKomnik}
and the Matsubara-voltage approach,\cite{Han}  have been applied. 
However, analytical approaches, 
 which should be 
complementary to the numerical ones, are still desired.

In this work, we study the high-energy limit, 
which is completely opposite to the ground state 
but is still non-trivial as the Coulomb repulsion $U$ 
remains competing with the hybridization energy scale $\Delta$. 
Furthermore, definitive knowledge of  
the opposite limit is helpful to clarify 
the nonequilibrium properties of 
the system for the whole energy regions.
Specifically, we consider the high bias limit $eV \to \infty$  
 where $eV$ is much greater than $T$ and 
 other energy scales of the impurity.
In this limit,  the model can be mapped exactly  
onto an effective two-site non-Hermitian Hamiltonian.
This can be deduced from an asymptotic form 
of the Keldysh Green's function,\cite{Keldysh,Caroli} using 
also a thermal-field-theoretical description.\cite{UmeMatTac,EzaAriHas}

We show that the dynamic susceptibilities for the charges,  spins, 
and the current noise,  do not depend on $U$ in 
the high bias limit. 
More generally, those correlation functions 
of the operators which commute with the total charge  
 take their noninteracting forms in the limit of $eV \to \infty$. 
Furthermore,
we show that the equation of motion for the Green's function 
 constitutes a closed system 
with two other equations of motion for 
the related higher-order correlation functions.
The exact solution in the high bias limit 
takes an extended form of 
the well-known 
atomic-limit solution,\cite{HubbardI,Donianch,HaugJauho} 
and depends sensitively on 
the asymmetry in the couplings,  $\Gamma_L$  and  $\Gamma_R$, 
between the impurity and two reservoirs on the left and right, 
respectively.
 This asymmetry, which is parameterized 
by $r \equiv (\Gamma_L-\Gamma_R)/\Delta$ 
with $\Delta \equiv \Gamma_L+\Gamma_R$,
 varies the impurity occupation from half-filling.
Furthermore, the self-energy captures a non-trivial 
imaginary part  of the value $3\Delta$ in the denominator. 
This value of the imaginary part coincides with 
 the corresponding term emerged 
in the order $U^2$ self-energy.\cite{AO2002}
 We also find that 
the order $U^2$  self-energy  becomes exact for $\Gamma_L = \Gamma_R$ 
 in the high bias limit.
A similar situation occurs also in the high-temperature limit $T \to \infty$ 
at thermal equilibrium $eV=0$ {\em without\/} 
the symmetry between  the  couplings $\Gamma_L$ and $\Gamma_R$.\cite{AO2002}
Namely, in the high-temperature limit,
 local correlation functions such as the impurity Green's function 
 do not depend on  $r$ and take the same functional forms  
as those in the symmetric-coupling  $\Gamma_L=\Gamma_R$  case 
of the high bias limit  $eV \to \infty$.

The thermal-field-theoretical approach,\cite{UmeMatTac,EzaAriHas} 
 which we will describe in a heuristic way,
is equivalent to the Keldysh formalism.
This approach may be regarded as a method of images applied to 
the Hilbert space, 
and similar formulation has recently 
been applied to quantum transport problems. 
\cite{EspositeHarolaMukamel,DzioevKosov,SaptosovWegewijs}
In this approach, the degrees of freedoms which 
correspond to the backward contour 
 in the Keldysh formalism are described by 
extra operators,  defined with respect to a doubled Hilbert space 
refereed to as Liouville-Fock space. 
By definition,  the extra operators 
for the enlarged part of the Hilbert space are independent 
of the original electron operators corresponding to the forward contour. 
However, a boundary condition is also imposed 
to the wavefunction at the turnaround point, $t= +\infty$, of 
the Keldysh contour in order to replicate the linear dependence 
among the four components of the  Keldysh correlation functions.

Our formulation is closely related to that   
of Saptosov and Wegewijs.\cite{SaptosovWegewijs} 
They have pointed out the solvability of 
the Anderson model in the high-temperature 
limit $T\to \infty$, 
where the average occupation of the impurity level is fixed at half-filling,  
 on the basis of 
 an observation that the  nonequilibrium reduced density matrix for 
the impurity site can be constructed exactly 
by a 16-component supervector with non-Hermitian superoperators 
defined in the Liouville-Fock space. 
They have studied in detail the structure of 
this Hilbert space in the high-temperature limit, 
by  classifying the basis set 
according to the symmetric properties of some conserved quantities, 
and have also demonstrated some numerical results for the symmetric 
couplings $\Gamma_L=\Gamma_R$.\cite{SaptosovWegewijs}
The 16-component basis apparently corresponds to 
the two-site degrees of freedom of fermions in our representation.
However, the physical back ground which leads 
some special properties to emerge in the high-energy limit,  
specifically properties of various correlation functions, 
still have not been fully clarified.

In this work, 
we consider the high bias limit $eV \to \infty$,  
 which  also has a correspondence 
to the high-temperature limit as mentioned.\cite{AO2002} 
In general case of the high bias limit, 
the asymmetry in the couplings $\Gamma_L \neq \Gamma_R$  
varies the average impurity occupation and  
affects significantly the correlation effects due to $U$.  
In order to clearly extract the underling physics,
we provide a heuristic derivation of 
the effective two-site Hamiltonian,  
starting from the path-integral representation 
of the time evolution along the Keldysh contour. 
The effective Hamiltonian is non-Hermitian, 
and can be expressed in terms of charges and currents. 
Furthermore, the equations of motion for 
 the charges and currents constitute a closed system, 
in a way that is somewhat similar to the case of 
the Tomonaga-Luttinger model.\cite{GL,ES} 
We describe these aspects    
in Sec.\ \ref{sec:Keldysh_formalism} and \ref{sec:effective_action}. 
Formulation for the correlation functions 
is given in Sec.\ \ref{sec:boundary_condition_TFD}.
Then, in Sec.\ \ref{sec:susceptibility},   
it is deduced from these equations of motion 
that the dynamic susceptibilities which conserve 
the total charge do not depend on $U$ in the high bias limit.
It is also the results of 
this charge and current dynamics 
 that the Green's function can be deduced exactly 
from a closed system of the three coupled equations of motion. 
The exact solution 
in the high bias limit is compared with the 
atomic-limit solution in Sec.\ \ref{sec:calculations_for_G}. 
Summary is given in Sec.\ \ref{sec:summary}.


\section{Keldysh formalism}
\label{sec:Keldysh_formalism}

We start with the single Anderson impurity 
coupled to two noninteracting leads; 
\begin{align}
\mathcal{H} =&   \sum_\sigma \epsilon_{d,\sigma}\, n_{d\sigma} 
 + U\,n_{d\uparrow}\,n_{d\downarrow} + 
\! \sum_{\alpha=L,R} \sum_{\sigma} 
\int_{-D}^D  \!\! d\epsilon\,  \epsilon\, 
 c^{\dagger}_{\epsilon \alpha \sigma} c_{\epsilon \alpha \sigma}^{}
\nonumber \\
& +   \sum_{\alpha=L,R} \sum_{\sigma}  v_{\alpha}^{}
 \left( \psi_{\alpha,\sigma}^\dag d_{\sigma}^{} + 
  d_{\sigma}^{\dag} \psi_{\alpha,\sigma}^{} \right) \;.
 \label{Hami_seri_part}
\end{align}
Here, $U$  is the Coulomb interaction between the electrons 
in the quantum dot,  and  $n_{d\sigma} = d^{\dag}_{\sigma} d^{}_{\sigma}$. 
 The operator $d^{\dag}_{\sigma}$ creates an electron 
in the quantized level of the energy $\epsilon _{d,\sigma}$  
which depends on spin $\sigma$ in the presence of a magnetic field.
 The operator $c_{\epsilon\alpha \sigma}^{\dagger}$ creates
an conduction electron in the lead 
on the left $\alpha=L$ or right $\alpha=R$ 
with   
$
\{ c^{\phantom{\dagger}}_{\epsilon\alpha\sigma'}, 
c^{\dagger}_{\epsilon'\alpha'\sigma'}
\} = \delta_{\alpha\alpha'} \,\delta_{\sigma\sigma'}   
\delta(\epsilon-\epsilon')$. 
The matrix element $v_{\alpha}^{}$  
couples the dot and the lead on $\alpha$ through 
$\psi^{}_{\alpha \sigma} \equiv  \int_{-D}^D d\epsilon \sqrt{\rho} 
\, c^{\phantom{\dagger}}_{\epsilon\alpha \sigma}$, 
and causes the level broadening 
 $\Delta \equiv \Gamma_L + \Gamma_R$ with   
 $\Gamma_{\alpha} = \pi \rho\, v_{\alpha}^2$ and $\rho=1/(2D)$. 
We consider the parameter region where 
the half band-width $D$  
 is much grater than the other energy scales, 
$D \gg \max(\Delta, U, |\epsilon_{d,\sigma}|, T, eV)$. 

\begin{figure}[tb]
 \leavevmode
\begin{minipage}{0.9\linewidth}
 \includegraphics[width=1\linewidth]{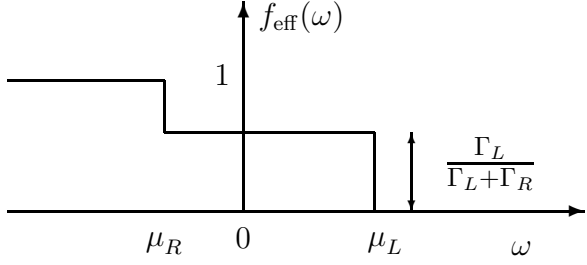}
\end{minipage}
%
%
%
%
%
%
\caption{
The nonequilibrium distribution function $f_{\mathrm{eff}}(\omega)$ 
for $\mu_L-\mu_R = eV$ and $T=0$.
The Fermi level at equilibrium, $eV=0$,
is chosen to be the origin of energy  $\omega=0$. 
}
\label{fig:distribution}
\end{figure}

The nonequilibrium steady state, driven by applied 
bias voltage, can be described  by an effective action 
$\mathcal{S} = \mathcal{S}_0+\mathcal{S}_U$  
for the time evolution in the Keldysh formalism;\cite{Keldysh,Caroli,MW,HDW}
\begin{align}
  &
  \!\!
   \mathcal{Z}  \,=\, 
  \int 
  \! D\overline{\eta}\, D\eta \ 
   e^{i \,\left[\,\mathcal{S}_0(\overline{\eta},\,\eta) \,+\,  
  \mathcal{S}_U(\overline{\eta},\,\eta)\,\right]}
  , \\
& 
\!\!
\mathcal{S}_0 \,=\,
   \sum_{\sigma} \int_{-\infty}^{\infty} dt\,dt'\; 
 \overline{\bm{\eta}}_{\sigma}^{}(t)\, 
        \bm{K}_{0,\sigma}(t, t')\, 
 \bm{\eta}_{\sigma}(t') \;, 
\label{eq:S0_Kldysh}
 \\
&  
\!\!\!
\mathcal{S}_U =
   -U \! \int \!\! dt \,
 \Bigl\{\,
  \overline{\eta}_{-,\uparrow}(t)\,
  \eta_{-,\uparrow}(t)\,
  \overline{\eta}_{-,\downarrow}(t)\,
\eta_{-,\downarrow}(t) 
\nonumber \\ 
 & \qquad \qquad \qquad  
 - \,   
  \overline{\eta}_{+,\uparrow}(t)\,
  \eta_{+,\uparrow}(t)\,
  \overline{\eta}_{+,\downarrow}(t)\,
  \eta_{+,\downarrow}(t) \Bigr\}.
\label{eq:SU_Kldysh}
\end{align}
Here, 
 $
 \overline{\bm{\eta}}_{\sigma}
 = \left(\, 
 \overline{\eta}_{-,\sigma}, \, 
 \overline{\eta}_{+,\sigma} \,\right) 
 $ is a pair of the Grassmann numbers for the $-$ and $+$ 
 branches of the Keldysh contour. 
The kernel  $\bm{K}_{0,\sigma}(t,t')$ 
is given by the Fourier transform of the noninteracting Green's function,
\begin{align}
& \ \quad 
\bm{K}_{0,\sigma}(t,t')  \, =   
      \int_{-\infty}^{\infty} \!{ d\omega \over 2 \pi }\,
      \left\{\bm{G}_{0,\sigma}(\omega)\right\}^{-1}
      e^{-i\omega (t-t')}  \;, 
 \label{eq:K0_Keldysh} \\ 
&
\!\!\!
\left\{\bm{G}_{0,\sigma}(\omega)\right\}^{-1} 
=  \,  (\omega-\epsilon_{d,\sigma}) 
\, \bm{\tau}_3^{} - \bm{\Sigma}_0(\omega) \;, 
\label{eq:G0_keldysh}
\\
& 
\qquad \    \bm{\Sigma}_0(\omega) \, = \,   
-i\Delta \bigl[\,1-2f_\mathrm{eff}(\omega) \,\bigr] 
\bigl( \bm{1} -  \bm{\tau}_1^{} \bigr)
+ \Delta \bm{\tau}_2^{} \;.
\label{eq:U0_self_keldysh}
\end{align}
Here,   $\bm{1}$ is the unit matrix and
  $\bm{\tau}_j$ for $j=1,2,3$ are the Pauli matrices,  
 \begin{align}
 \!\!\!
 \bm{\tau}_1 =
 \left(
 \begin{matrix}
 0 &   1 \cr
 1 &  0 \cr  
 \end{matrix}
 \right) , 
 \quad
 \bm{\tau}_2 =
 \left(
 \begin{matrix}
 0 &   -i \cr
 i &  \ 0  \cr  
 \end{matrix}
 \right) , 
 \quad
 \bm{\tau}_3 =
 \left(
 \begin{matrix}
 1 &  \  0 \cr
 0 &  -1 \cr  
 \end{matrix}
 \right) .
 \end{align}
The distribution function 
 $f_{\mathrm{eff}}(\omega)$ describes the  energy window 
as shown in Fig.\ \ref{fig:distribution}, and is defined by 
\begin{equation}
f_{\mathrm{eff}}(\omega) 
\, = \, 
\frac{\Gamma_L\,f_L(\omega) + \Gamma_R\,f_R(\omega)}
 { \Gamma_L +\Gamma_R } \;.
\label{eq:f_eff}
\end{equation}
Here,  $f_{\alpha}(\omega) 
= [\,e^{(\omega-\mu_\alpha)/T}+1\,]^{-1}$ and
 $\mu_\alpha$ is the chemical potential for lead $\alpha$.
This function  $f_\mathrm{eff}(\omega)$ 
also  determines the long time behavior 
of $\bm{K}_{0,\sigma}(t,t')$ as 
a function of $t-t'$.
Furthermore, 
the temperature $T$ and 
bias voltage $eV \equiv \mu_L - \mu_R$ 
 enter through  $f_\mathrm{eff}(\omega)$ 
for the local correlation functions for the impurity site.

\section{Liouville-Fock space for $eV \to \infty$}
\label{sec:effective_action}

We now consider two kinds of the high-energy limits.
One is the high bias limit $eV\gg T$, 
where $f_L \to 1$ and $f_R \to 0$. 
The other is high-temperature limit $T \gg eV$, 
where $f_L=f_R \to 1/2$, including thermal equilibrium at $eV=0$ 
as a special case.
In both of the limits, 
the distribution function $f_\mathrm{eff}(\omega)$ becomes 
a constant independent of the frequency $\omega$,
\begin{align}
\!\!\!\!\!\!\!\!
f_\mathrm{eff}(\omega) 
\, \to \left\{  
\begin{array}{lr}
 \frac{\Gamma_L}{\Gamma_L+\Gamma_R} \;, &   \mbox{for} \ eV \to \infty 
\\
\quad \frac{1}{2}  \quad \ , & \mbox{for} \ \ \,  T \to \infty  
\rule{0cm}{0.4cm}
\end{array}
\right. \; ,
\label{eq:f_eff_high_energy}
\end{align}
and then the non-interacting self-energy  $\bm{\Sigma}_0(\omega)$ 
also becomes independent of $\omega$.
This makes the problems in the high-energy limits solvable. 
In the following, we concentrate on the $eV \to \infty$ limit 
because the $T \to \infty$ limit is equivalent to 
the $\Gamma_L =\Gamma_R$ case of the $eV \to \infty$ limit 
as long as the local properties at the impurity site are concerned.

\subsection{Effective non-Hermitian Hamiltonian}

 In this high bias limit, 
Eq.\ $\eqref{eq:U0_self_keldysh}$ simplifies in the form 
\begin{align}
& 
\lim_{eV \to \infty} \!  \bm{\Sigma}_0(\omega) 
\,=\, 
\bm{\tau}_3 \,\bm{L}_0^{} \;, 
\label{eq:L0_matrix_Vinf_tau3}
\\
& \!\!\!
\bm{L}_0^{} 
\, \equiv  \, 
i\left [ \,
\begin{matrix} 
 \Gamma_L - \Gamma_R   & -2\Gamma_L\,  \cr
 -2\Gamma_R  &  -(\Gamma_L - \Gamma_R)      
\rule{0cm}{0.5cm}  
\cr
\end{matrix}          
\right]  . 
\label{eq:L0_matrix_Vinf}
\end{align}
In this limit, the kernel $\bm{K}_{0,\sigma}(t,t')$ 
 becomes a short-time function of a Markovian form 
and vanishes for $t \neq t'$. 
The explicit form contains the $t$ dependence only through 
the first derivative 
which arises from the $\omega$ 
linear part of $\left\{\bm{G}_{0,\sigma}(\omega)\right\}^{-1}$,  
\begin{align}
& 
\! 
   \mathcal{S}_0 \to 
   \sum_{\sigma} \! \int_{-\infty}^{\infty} \!\! dt \,  
\bm{\eta}_{\sigma}^{\dagger}(t) 
\left\{
\bm{1}  
\!\left(\! i \frac{\partial}{\partial t}  - \epsilon_{d,\sigma} \!\right)\! 
- \bm{L}_0   
\right\} 
 \bm{\eta}_{\sigma}(t) . 
  \label{eq:S0_Vinf} 
\end{align}
Here, 
a transformation, 
$\bm{\eta}_{\sigma}^{\dagger} =  \overline{\bm \eta}_{\sigma}\,\bm{\tau}_3$  
keeping the counter part ${\bm \eta}_{\sigma}$ unchanged,  
has been introduced 
in order to take the time-derivative term diagonal.
The interaction part of action $\mathcal{S}_U$ does not change 
the form  of Eq.\ \eqref{eq:SU_Kldysh} 
under this transformation,
\begin{align}
& 
\mathcal{S}_U \,= \, 
   -U \! \int \!\! dt \,
 \Bigl\{\,
  \eta_{-,\uparrow}^{\dagger}(t)\,
  \eta_{-,\uparrow}(t)\,
  \eta_{-,\downarrow}^{\dagger}(t)\,
  \eta_{-,\downarrow}(t) 
\nonumber \\ 
 & \qquad   \qquad \qquad \ \ 
-  \,   
  \eta_{+,\uparrow}^{\dagger}(t)\,
  \eta_{+,\uparrow}(t)\,
  \eta_{+,\downarrow}^{\dagger}(t)\,
  \eta_{+,\downarrow}(t) \Bigr\}.
\label{eq:SU_TFD}
\end{align}
Therefore, in the high bias limit, the integrand for 
the effective action $\mathcal{S}_0 + \mathcal{S}_U$, 
namely the Lagrangian, does not have an explicit time-dependence   
other than 
the first derivative $i\partial/\partial t$ in Eq.\  \eqref{eq:S0_Vinf}.

The effective action of this form can be constructed from 
a non-Hermitian Hamiltonian which consists of two orbitals;  
 $\,\widehat{H}_\mathrm{eff}^{} = \widehat{H}_\mathrm{eff}^{(0)} 
 + \widehat{H}_\mathrm{eff}^{(U)}$,
\begin{align}
 \widehat{H}_\mathrm{eff}^{(0)} =  & \ 
\sum_\sigma \xi_{d,\sigma} \,Q_{\sigma}^{} 
\,+ 
\sum_\sigma 
\left( 
\bm{d}_{\sigma}^{\dagger} \bm{L}_0^{} \bm{d}_{\sigma}^{} 
 - i\Delta 
\right) , 
\label{eq:H_0_Vinf}
\\
 \widehat{H}_\mathrm{eff}^{(U)}   =&  
\  U \biggl[
  \Bigl(\!n_{-,\uparrow }^{} \! - \! \frac{1}{2}\Bigr) 
\!
  \Bigl(\!n_{-,\downarrow }^{} \! - \!  \frac{1}{2}\Bigr) 
 \! - \!    \Bigl(\!n_{+,\uparrow }^{} \! - \!  \frac{1}{2}\Bigr)
\!  
  \Bigl(\!n_{+,\downarrow }^{} \! - \!  \frac{1}{2}\Bigr)  
 \biggr]
\nonumber \\
= & \    
U \left( 
\,Q_{\uparrow}^{} q_{\downarrow}^{} 
+ Q_{\downarrow}^{} q_{\uparrow}^{} 
\right) .  
\label{eq:H_U_Vinf}
\end{align}
Here, 
 $\xi_{d,\sigma} = \epsilon_{d,\sigma} +U/2$,  and  
$
 {\bm{d}}_{\sigma}^{\dagger}
 = 
\bigl( d_{-,\sigma}^{\dagger}\,, \,  d_{+,\sigma}^{\dagger}\bigr) 
$ is a set of two independent fermion operators 
introduced for the  $-$ and $+$ branches, respectively. 
The corresponding charges are defined by 
$n_{\mu,\sigma} =  d_{\mu,\sigma}^{\dagger} d_{\mu,\sigma}^{}$,   
\begin{align}
q_{\sigma}^{} \equiv \frac{n_{-,\sigma} - n_{+,\sigma}}{2} ,
\qquad \   
Q_{\sigma}^{} \equiv   n_{-,\sigma} + n_{+,\sigma} -1, 
\label{eq:charge_q_Q}
\end{align}
and $Q \equiv \sum_\sigma Q_{\sigma}$ is  the total charge.
In this representation, 
the fermion operators with the label \lq\lq $-$" describe the original 
impurity electron $d_{-,\sigma}^{\dagger} \equiv d_{\sigma}^{\dagger}$. 
The other component with the label \lq\lq $+$" corresponds to 
a tilde-conjugate operators $\widetilde{d}_{\sigma}^{\dagger}$ 
in the standard notation 
of the thermal field theory\cite{UmeMatTac,EzaAriHas} 
and is defined in a doubled Hilbert space,  
which is also referred to as a Liouville-Fock space.
\cite{EspositeHarolaMukamel,DzioevKosov,SaptosovWegewijs}
Specifically, our representation corresponds to 
a particle-hole transformed version, in which  
 $\,d_{+,\sigma}^{\dagger} \equiv \widetilde{d}_{\sigma}^{}$.

The time evolution in the doubled Hilbert space 
can be described  by the Heisenberg representation 
defined with respect to 
$\widehat{H}_{\mathrm{eff}}$, \cite{UmeMatTac,EzaAriHas}  
\begin{align}
\mathcal{O}(t) \,\equiv& \  
e^{i\widehat{H}_{\mathrm{eff}} t}
\,\mathcal{O}\,
e^{-i\widehat{H}_{\mathrm{eff}} t} \;,
\\
i \,\frac{\partial \mathcal{O}(t)}{\partial t} \,= & \  
\left[\,\mathcal{O}(t)\,,\,\widehat{H}_{\mathrm{eff}} \,\right] 
\;.
\end{align}

\subsection{Charge and current representation} 
\label{subseec:current_algebra}

The Liouville-Fock space for $\widehat{H}_{\mathrm{eff}}$
 consists of $4^2$ states, which   
can be classified according to the total 
charge  and spin.
One of the meaningful observations on this is 
that $\widehat{H}_\mathrm{eff}^{}$ 
can be expressed in terms of the charges and currents, 
\begin{align}
\widehat{H}_\mathrm{eff}^{}
 \,= & \    
\sum_\sigma \xi_{d,\sigma} \,Q_{\sigma}^{} 
+i \sum_\sigma \left( P_{\sigma}^{} -\Delta \right) , 
\label{eq:H_Vinf_with_currents}
\\
 P_{\sigma}^{} \,\equiv & \     
 I_{R,\sigma} + I_{L,\sigma} 
 + 
 2\, \mathcal{W}_{\sigma}^{}\, q_{\sigma}^{} , 
\label{eq:current_P}
 \\
 \mathcal{W}_{\sigma}^{} \, \equiv& \        
  (\Gamma_L -\Gamma_R) - i\, \frac{U}{2}  \,Q_{\overline{\sigma}}^{} \;.
\label{eq:W_overline} 
\end{align}
The Coulomb interaction enters through  the last term 
 in the right-hand-side of Eq.\ \eqref{eq:W_overline}, 
where the label $\overline{\sigma}$  for  $Q_{\overline{\sigma}}$ 
denotes the spin opposite to $\sigma$.
The  operators  $I_{R,\sigma}^{}$ and  $I_{L,\sigma}^{}$ are  defined by
\begin{align}
 &I_{R,\sigma}^{} =  
 -2 \Gamma_R \,d_{+,\sigma}^{\dagger}d_{-,\sigma}^{} ,
  \quad \ 
 I_{L,\sigma}^{} =  
 -2 \Gamma_L \,d_{-,\sigma}^{\dagger}d_{+,\sigma}^{}  .
 \label{eq:current_eV_inf_def} 
 \end{align}
These operators are equivalent 
to the current flowing from the dot to the right lead 
and the current flowing from the left lead to the dot, respectively,   
although they are not Hermitian in the Liouville-Fock space.
We find  that  $Q_{\sigma}$ and $P_{\sigma}$ are conserved, and 
the equations of motion  for  the relative charge $q_{\sigma}^{}$ 
defined in Eq.\ \eqref{eq:charge_q_Q}
and the relative current 
\begin{align}
p_{\sigma}^{} \,\equiv\,  I_{R,\sigma} - I_{L,\sigma}  
\end{align}
consititute a closed system,
\begin{align}
& 
\frac{\partial Q_{\sigma}^{}}{\partial t} =   0  ,
\qquad  \quad 
\frac{\partial P_{\sigma}^{}}{\partial t}  = 0, 
\label{eq:EOM_Qtot_Vinf}
\\ 
& 
 \frac{\partial q_{\sigma}^{}}{\partial t} = - \, p_{\sigma}^{} ,  \qquad 
\frac{\partial p_{\sigma}^{}}{\partial t} 
= \, 
 4\, \mathcal{L}_{\sigma}^{2} \,q_{\sigma}^{}
+ 
 2\,\mathcal{W}_{\sigma}^{} P_{\sigma}^{} \;.  
\label{eq:EOM_Qtot_Vinf_2}
\end{align}
Here, the operators 
 $\mathcal{L}_{\sigma}^{2}$ and 
$\mathcal{W}_{\sigma}^{}$ take 
complex eigenvalues which    
depend on the conserved charge $Q_{\overline{\sigma}}$  as 
\begin{align}
\mathcal{L}_{\sigma}^{2}  \,\equiv  \, 
\left(\frac{U}{2}\right)^2 Q_{\overline{\sigma}}^2
\,+\, i\, (\Gamma_L-\Gamma_R) \, U  Q_{\overline{\sigma}}^{}
\,-\,  \Delta^2   .
 \label{eq:L2_effective_def}
\end{align}
Furthermore, 
it is deduced from Eqs.\ \eqref{eq:EOM_Qtot_Vinf} and 
\eqref{eq:EOM_Qtot_Vinf_2}  that 
the second derivative of  $p_{\sigma}$ takes a simple form 
\begin{align}
\frac{\partial^2 p_{\sigma}^{}}{\partial t^2} \,= & \ 
- 4\, \mathcal{L}_{\sigma}^{2} \,p_{\sigma}^{} \;.
\label{eq:EOM_IL_Vinf}
\end{align}

Specifically, in the subspace of  $Q_{\overline{\sigma}}=0$, 
the operator $\mathcal{L}_{\sigma}^{2}$ takes the eigenvalue of $-\Delta^2$ 
that does not depend  on the Coulomb interaction. 
Therefore, 
the Heisenberg operators for 
the charges and currents can be expressed simply 
 as a linear combination 
of $e^{2 \Delta t}$ and $e^{-2 \Delta t}$, 
which represent relaxation of a particle-hole pair excitation.  
Furthermore, due to these properties, 
 the susceptibilities for the charges and currents 
become independent of the Coulomb interaction  
in the high bias limit and take the noninteracting forms 
as it will be discussed in Sec.\ \ref{sec:susceptibility}.
We will also show in Sec.\ \ref{sec:calculations_for_G} that 
the equations of motion given in Eqs.\ \eqref{eq:EOM_Qtot_Vinf} and 
\eqref{eq:EOM_Qtot_Vinf_2} also 
determine essential dynamics of the 
single-particle Green's function as well as 
the fluctuations of the charges and currents.

\section{Two reference states:  
$\langle \! \langle I |\! |$ and  $|\! | \rho \rangle \! \rangle$ 
}
\label{sec:boundary_condition_TFD}

In this section, 
we introduce the explicit forms of the final and initial states  
for the time-dependent perturbation theory  
 in the doubled Hilbert space.  
These two are also referred to as  
 $\langle \! \langle I |\! |$ and  $|\! | \rho \rangle \! \rangle$ 
in the standard notation 
of the thermal field theory,\cite{UmeMatTac,EzaAriHas} or   
 the Liouville-Fock 
approaches.\cite{EspositeHarolaMukamel,DzioevKosov,SaptosovWegewijs} 
Specifically, $\langle \! \langle I |\! |$ is defined 
such that it satisfies the boundary conditions at $t=+\infty$ 
while $|\! | \rho \rangle \! \rangle$ includes the information 
about the statistical distribution.
We describe the properties of these two states in the following.

\subsection{Time evolution in the interaction representation}
\label{subsec:interaction_representation}

In this subsection,  
we discuss more in detail the time evolution,
which is described in the interaction representation  
by the operators,   
\begin{align}
\widehat{\mathcal{U}}(t_2,t_1) 
 \equiv& \  \mathrm{T} \exp  \left[ 
-i    \displaystyle 
\int_{t_1}^{t_2}     dt \, 
e^{i\widehat{H}_\mathrm{eff}^{(0)} t} 
\widehat{H}_\mathrm{eff}^{(U)}
e^{-i\widehat{H}_\mathrm{eff}^{(0)} t}
\right] , 
\label{eq:U_time_evolution} \\
\mathcal{O}^{\mathcal{I}}_{}(t) \,\equiv & \ \,  
  e^{i\widehat{H}_\mathrm{eff}^{(0)} t}\, 
  \mathcal{O}\,
  e^{-i\widehat{H}_\mathrm{eff}^{(0)} t} \;.
\end{align}
Here, $\mathrm{T}$ in the right-hand side of Eq.\ \eqref{eq:U_time_evolution}
is the time-ordering operator.

The unperturbed part of the effective Hamiltonian 
can be rewritten in a diagonal form 
\begin{align} 
\widehat{H}_\mathrm{eff}^{(0)} 
\,=& \ 
\sum_\sigma  \xi_{d,\sigma}
\left( 
a_{\sigma}^{-1} a_{\sigma}^{} 
+
b_{\sigma}^{-1} b_{\sigma}^{} 
 -1  
\right) 
\nonumber \\
& \ +\,  
\sum_\sigma \, i \Delta 
\left( 
a_{\sigma}^{-1} a_{\sigma}^{} 
-
b_{\sigma}^{-1} b_{\sigma}^{}
-1 \right)  \;.
\end{align} 
Here, the operators correspond to the left and right eigenvectors 
of $\bm{L}_0^{}$, 
\begin{align} 
&\!\!\!
a_{\sigma}^{} 
 \equiv    
\frac{ d_{-,\sigma}^{} - d_{+,\sigma}^{}
}{\sqrt{2}}
,  \quad
a_{\sigma}^{-1} 
 \equiv 
\frac{\sqrt{2}\bigl(\Gamma_L d_{-,\sigma}^{\dagger}
\! - \Gamma_R d_{+,\sigma}^{\dagger}\bigr)}{\Gamma_L+\Gamma_R},   
\label{eq:Vinf_gamma_operators_1}
\\
& \!\!\!
b_{\sigma}^{-1} 
 \equiv    
\frac{d_{-,\sigma}^{\dagger} +  \, d_{+,\sigma}^{\dagger}
}{\sqrt{2}} , 
\quad 
b_{\sigma}^{} 
 \equiv 
\frac{\sqrt{2}\bigl(
\Gamma_R d_{-,\sigma}^{\phantom{|}}
\! +  \Gamma_L d_{+,\sigma}^{\phantom{|}} \bigr)}{\Gamma_L+\Gamma_R} .    
\label{eq:Vinf_gamma_operators_2}
\end{align} 
 These operators satisfy the anti-commutation relations;   
\begin{align}
& 
\!\!
\bigl\{ a_{\sigma}^{} , a_{\sigma'}^{-1} \bigr\}  =  
\bigl\{ b_{\sigma}^{} , b_{\sigma'}^{-1} \bigr\} = \, \delta_{\sigma\sigma'} 
\;, \\
& 
\!\!
\bigl\{ b_{\sigma}^{} , a_{\sigma'}^{-1} \bigr\} = 
\bigl\{ a_{\sigma}^{} , b_{\sigma'}^{-1} \bigr\} = 
\bigl\{ a_{\sigma}^{} , a_{\sigma'\! } \bigr\} = 
\bigl\{ b_{\sigma}^{} , b_{\sigma'\! } \bigr\} = 0.
\end{align}

These operators show different types of long-time behavior 
in the interaction representation,
\begin{align}
 a_{\sigma}^{\mathcal{I}}(t) =  
 a_{\sigma}^{} e^{(\Delta -i\xi_{d,\sigma}) t},
\quad  \ 
 b_{\sigma}^{\mathcal{I}}(t) =
 b_{\sigma}^{} e^{-(\Delta+i\xi_{d,\sigma}) t}  .  
\label{eq:a_b_interaction_rep}
\end{align}
This shows that the annihilation operators decay or expand in time 
depending on the imaginary part of the eigenvalue 
because of the non-Hermiticy of $\widehat{H}_\mathrm{eff}^{(0)}$. 
Therefore, 
the final and initial states 
 satisfy a strong requirement that 
they  must describe the relaxation of the correlation functions correctly. 
We find that such states 
can be constructed  from 
the left and right eigenstates of doubly-occupied $a_{\sigma}^{}$ particles, 
\begin{align}
\langle \! \langle I |\! |  \,\equiv \, 
\left\langle 0 \right| 
a_{\downarrow}^{} a_{\uparrow}^{} \;, 
\qquad 
|\! |  \rho  \rangle \! \rangle 
\,\equiv\, 
a_{\uparrow}^{-1} 
a_{\downarrow}^{-1} 
\left|0\right\rangle  \;.
\label{eq:initial_final_states}
\end{align}
One of the reasons is that the components showing an exponential growth 
can be eliminated only in the case where 
the propagators are defined with respect to this pair of the states,
as  
\begin{align} 
\langle\!\langle I|\!|
\mathrm{T}\,
a^{\mathcal{I}}_{\sigma}(t)\, a^{-1\mathcal{I}}_{\sigma'}(0)
|\! | \rho \rangle \! \rangle 
\,=& \, - \delta_{\sigma\sigma'}\, \theta(-t)\,e^{(\Delta -i\xi_{d,\sigma}) t} 
,
\label{eq:G0_aa_Vinf}
\\
\langle\!\langle I|\!|
\mathrm{T}\,
b^{\mathcal{I}}_{\sigma}(t)\, b^{-1\mathcal{I}}_{\sigma'}(0)
|\! | \rho \rangle \! \rangle 
\,= &  \ \ \  
\delta_{\sigma\sigma'} \,\theta(t)\, e^{-(\Delta+i\xi_{d,\sigma}) t} 
\;.
\label{eq:G0_bb_Vinf}
\end{align}
As a result, these propagators decay for $|t|\to \infty$  properly. 
We consider the interacting propagators  in Sec.\ \ref{subsec:TFD_green}.

The final state $\langle \! \langle I |\! |$, 
defined in Eq.\ \eqref{eq:initial_final_states},
also satisfies the following relations, 
\begin{align}
\langle \! \langle I |\! | d_{-,\sigma}^{}
 =  \langle \! \langle I |\! | d_{+,\sigma}^{}\;, \qquad
\langle \! \langle I |\! |  d_{-,\sigma}^{\dagger} 
= - \langle \! \langle I |\! |   d_{+,\sigma}^{\dagger} \;.
\label{eq:bra_boundary}
\end{align}
These relations are necessary for the construction of the approach 
and can be regarded as the boundary conditions 
which replicate the mutual dependence between the $-$ and $+$ components 
of the Keldysh contour.
Furthermore, in the high bias limit, 
 the initial state $|\!| \rho \rangle\!\rangle$  also  
has similar properties,  
\begin{align}
 \Gamma_L\, d_{+,\sigma}^{} |\!| \rho \rangle\!\rangle\, =&\ 
-\Gamma_R\, d_{-,\sigma}^{} |\!| \rho \rangle\!\rangle\;,  
\label{eq:ket_boundary_1} \\
 \Gamma_R \,d_{+,\sigma}^{\dagger} |\!| \rho \rangle\!\rangle \,= & \ \ 
\Gamma_L\, d_{-,\sigma}^{\dagger} |\!| \rho \rangle\!\rangle \;. 
\label{eq:ket_boundary_2}
\end{align}
We see in these relations  
 that the Hamiltonian parameters 
$\Gamma_L$ and $\Gamma_R$ appear as the coefficients. 
These reflect the properties of 
the density matrix in the limit of $eV \to \infty$, 
described in the following.

\subsection{Density matrix}
\label{subsec:density_matrix_Vinf}

The two states 
 $\langle \! \langle I |\! |$ and  $|\! | \rho \rangle \! \rangle$,   
defined in Eq.\ \eqref{eq:initial_final_states},
 have some other notable properties:  
both of them belong to a spin-singlet subspace 
with the total charge $Q=0$ 
and are also the eigenstates of  $P_{\sigma}$, as 
\begin{align}
&\langle\!\langle I|\!| Q_{\sigma} =\, 0 \;,
\qquad \qquad 
\langle\!\langle I|\!| P_{\sigma} \,=\, 
\langle\!\langle I|\!|\,\Delta
\; ,   \\
& 
Q_{\sigma} |\!| \rho \rangle \! \rangle \,=\, 0 \;, 
\qquad \qquad  
 P_{\sigma} |\!| \rho \rangle \! \rangle 
\,=\, \Delta\,|\!| \rho \rangle \! \rangle \;.
\label{eq:Q=0} 
\end{align}
Furthermore, 
  $\langle\!\langle I|\!|$ and  $|\!| \rho\rangle \! \rangle$ 
are the zero-energy eigenstates of both   
$\widehat{H}_\mathrm{eff}^{(0)}$  and  $\widehat{H}_\mathrm{eff}^{}$,    
\begin{align}
& 
\langle\!\langle I|\!| \widehat{H}_\mathrm{eff}^{(0)}= \,0\;,
  \qquad 
\langle\!\langle I|\!| \widehat{H}_\mathrm{eff}^{}\,=\,0\;, 
\label{eq:zero_mode_H_eff_bra}
\\
&
\widehat{H}_\mathrm{eff}^{(0)} |\!|\rho \rangle \! \rangle = \,0\;,  
\qquad \, 
 \widehat{H}_\mathrm{eff}^{} |\!|\rho \rangle \! \rangle \,= \,0\;.   
\label{eq:zero_mode_H_eff_ket}
\end{align}
Due to these zero values of the eigenenergies, these two states 
 do not evolve in time, 
\begin{align}
& \!\!\!\!\! 
\langle\!\langle I|\!|\,\widehat{\mathcal{U}}(t,t')  
= \langle\!\langle I|\!|, \quad \  \, 
|\! | \rho(t) \rangle \! \rangle 
 \equiv 
\widehat{\mathcal{U}}(t, -\infty) 
|\! | \rho \rangle \! \rangle  =  
|\! | \rho \rangle \! \rangle  . \!\!\!  
\label{eq:initial_final_no_evolve}
\end{align}

We consider statistical averages 
which  are defined with respect to     
 $\langle\!\langle I|\!|$ and  $|\!| \rho(t)\rangle \! \rangle$  at $t=0$,  
\begin{align}
\langle \mathcal{O}(t) \rangle \equiv 
\langle\!\langle I|\!| \mathcal{O} (t)|\!| \rho(0) \rangle \! \rangle
\ = \ 
\langle\!\langle I|\!| \mathcal{O}(t) |\!| \rho\rangle \! \rangle \;.
\label{eq:average_Vinf}
\end{align}
Here, the second equation has been obtained 
by using Eq.\ \eqref{eq:initial_final_no_evolve}. 
We note that  the normalization 
condition  $\langle\!\langle I|\!| \rho \rangle \! \rangle = 1$
is satisfied by definition, given in Eq.\ \eqref{eq:initial_final_states}. 
The underlying statistical weight can be extracted 
as an Hermitian density matrix, which satisfies    
$
\widehat{\rho}\,|\! |  I  \rangle \! \rangle =
|\! |  \rho  \rangle \! \rangle$, 
\begin{align}
  \widehat{\rho} \, = & \, 
 \prod_\sigma \bigl(\, 1 +  
2 r \, q_{\sigma}
 \bigr) \;, 
\qquad  \ \ 
r \equiv \frac{\Gamma_L-\Gamma_R}{\Gamma_L+\Gamma_R} \;. 
\label{eq:density_matrix_Vinf}
\end{align}
This density matrix 
for the high bias limit has the properties that 
the statistical weight does not depend on $U$ 
but varies as 
a function of $r$ which parameterizes the asymmetry 
in the couplings. Specifically, for the symmetric couplings  $r = 0$,   
the distribution becomes uniform  $\widehat{\rho} = 1$.

The average formula, Eq.\  \eqref{eq:average_Vinf}, 
reproduces the exact value of the number of the electrons 
occupied in the dot in the high bias limit 
[see Appendix \ref{sec:some_expectation_values}].
\begin{align} 
\langle\!\langle I|\!| n_{-,\sigma} |\!|\rho(0) \rangle \! \rangle 
\,= \, \frac{\Gamma_L}{\Gamma_L+\Gamma_R}\;,
\end{align}
and  $\langle n_{+,\sigma} \rangle = 1- \langle  n_{-,\sigma}  \rangle$.  
Furthermore, the steady currents through 
the dot are correctly reproduced and  are properly conserved,      
\begin{align} 
 &
 \!\!\!\!
 \langle\!\langle I|\!| I_{R,\sigma} |\!|\rho(0) \rangle \! \rangle 
  =  
 \langle\!\langle I|\!| I_{L,\sigma} |\!|\rho(0) \rangle \! \rangle 
 \ = \   
 \frac{2\Gamma_L\Gamma_R}{\Gamma_L+\Gamma_R}\;.
 \end{align}
Even in the presence of the Coulomb repulsion $U$,  
the averages of the charges and currents really take the same values 
as those in the noninteracting case  in the high bias limit,
as shown in the Appendix \ref{sec:some_expectation_values}. 
The same holds also true for a class of correlation functions, 
such as the dynamic susceptibilities of charge and currents, 
 as shown in Sec.\ \ref{sec:susceptibility}.

\subsection{Green's function for the doubled Hilbert space}
\label{subsec:TFD_green}


In this subsection, we describe the relation 
between the Green's function defined 
with respect to the Liouville-Fock space and 
the original Keldysh Green's function.
We introduce the noninteracting Green's function, defined by 
\begin{align} 
 \mathcal{G}_{0,\sigma}^{\mu\nu}(t) 
\, \equiv  & \  - i \,
\langle\!\langle I|\!|
\mathrm{T}\,
d^{\mathcal{I}}_{\mu,\sigma}(t)\, d^{\dagger\mathcal{I}}_{\nu,\sigma}(0)
|\! | \rho \rangle \! \rangle \;.
\label{eq:G_eff_Vinf_0}
\end{align}
This Green's function can be calculated by  
using Eqs.\ \eqref{eq:a_b_interaction_rep}
and \eqref{eq:initial_final_states}. 
The result 
shows a one-to-one correspondence 
\begin{align}
& \bm{\mathcal{G}}_{0,\sigma}^{}(t)  
  \equiv   
 \left[\, 
  \begin{matrix}
   \mathcal{G}^{--}_{0}(t) & \mathcal{G}^{-+}_{0}(t)   \cr
   \mathcal{G}^{+-}_{0}(t) & \mathcal{G}^{++}_{0}(t)  \cr  
  \end{matrix} 
  \, \right]  
\  = \   
\bm{G}_{0,\sigma}^{}(t) \, \bm{\tau}_3
 \label{eq:G0_TFD_Vinf} 
\end{align}
with the Keldysh Green's function  $\bm{G}_{0,\sigma}^{}$,
the explicit form of which in the high bias limit  
is given by substituting Eq.\ \eqref{eq:L0_matrix_Vinf_tau3} 
into Eq.\ \eqref{eq:G0_keldysh} and 
replacing 
$\epsilon_{d,\sigma}$ by $\xi_{d,\sigma}$, following   
the definition of $\widehat{H}_\mathrm{eff}^{(0)}$, as 
 \begin{align}
 & \left\{\bm{G}_{0,\sigma}(\omega)\right\}^{-1} 
 = \, \bm{\tau}_3^{} \,\Bigl[\,  
(\omega-\xi_{d,\sigma}) \bm{1} 
 \, - \,\bm{L}_0 \, \Bigr]\;.
 \label{eq:G0_keldysh_rev} 
\end{align}
For instance, the retarded Green's function is given by  
 $G_{0,\sigma}^{r}(t)= 
-i\theta(t) \, e^{-(\Delta+i\xi_{d,\sigma}) t}$ in real time. 
This explicitly indicates that  
the dynamics with the non-Hermitian relaxations can be described 
correctly by the final and initial states 
defined in Eq.\ \eqref{eq:initial_final_states}.

For interacting case with $U\neq 0$, 
the full Green's function in the Liouville-Fock space is defined by 
\begin{align} 
 \mathcal{G}_{\sigma}^{\mu\nu}(t) 
\, \equiv  & \  - i \,
\langle\!\langle I|\!|
\mathrm{T}\,
d^{\phantom{\dagger}}_{\mu,\sigma}(t)\, d^{\dagger}_{\nu,\sigma}(0)
|\! | \rho(0) \rangle \! \rangle 
\label{eq:G_eff_Vinf}
\\
 =  & \,  - i \,
\langle\!\langle I|\!|
\mathrm{T}\,
d^{\mathcal{I}}_{\mu,\sigma}(t)\, d^{\dagger\mathcal{I}}_{\nu,\sigma}(0)
\,\widehat{\mathcal{U}}(\infty, -\infty) 
|\! | \rho \rangle \! \rangle \;.
\label{eq:G_eff_Vinf_interaction_representation}
\end{align}
The same relation between 
$\bm{\mathcal{G}}_{\sigma}^{}$ and 
$\bm{G}_{\sigma}^{}$, which is the interacting   
Keldysh Green's function, as that in the noninteracting case 
also holds for $U\neq 0$,
\begin{align}
\bm{G}_{\sigma}^{}(t) 
\,=& \  
\bm{\mathcal{G}}_{\sigma}^{} (t) \, \bm{\tau}_3 
\;. 
\label{eq:GKldysh_Gtfd}
\end{align}
The Feynman diagrammatic expansion based on 
the Wick's theorem is applicable to 
the interaction representation, 
Eq.\ \eqref{eq:G_eff_Vinf_interaction_representation}, 
with the final and initial states 
defined in Eq.\ \eqref{eq:initial_final_states}.
The diagrams generated for $\bm{\mathcal{G}}_{\sigma}^{}$  
have exact one-to-one correspondence to those 
for $\bm{G}_{\sigma}^{}$ obtained with the Keldysh perturbation expansion 
 [see Appendix \ref{sec:Hartree}].

Using the linear dependence among the  four components of 
the Keldysh Green's function,  the retarded and advanced functions 
can be expressed in the form 
\begin{align}
G_{\sigma}^r 
\,=& \  
\mathcal{G}_{\sigma}^{--} + \mathcal{G}_{\sigma}^{-+}
\,=\, 
\mathcal{G}_{\sigma}^{+-} + \mathcal{G}_{\sigma}^{++} \;, 
\label{eq:Gr_Keldysh_sum_rule}
\\
G_{\sigma}^a 
\,=& \  
\mathcal{G}_{\sigma}^{--} - \mathcal{G}_{\sigma}^{+-}
\,=\, 
\mathcal{G}_{\sigma}^{++} - \mathcal{G}_{\sigma}^{-+} \;.
\label{eq:Ga_Keldysh_sum_rule}
\end{align}
In the right-hand side of these two equations, 
the second equalities for $\mathcal{G}_{\sigma}^{\mu\nu}(t)$'s 
can also be deduced directly from Eq.\ \eqref{eq:G_eff_Vinf} 
using Eq.\ \eqref{eq:bra_boundary}.

It can also  be deduced 
from  Eqs.\ \eqref{eq:bra_boundary}-\eqref{eq:ket_boundary_2} 
that each of the four components of $\mathcal{G}_{\sigma}^{\mu\nu}$ 
can be expressed in terms of  $G_{\sigma}^r$ 
and  $G_{\sigma}^a$  in the limit $eV \to \infty$, 
\begin{align}
\!\!
\bm{\mathcal{G}}_{\sigma}^{}(t)
\,=\,    
 \frac{G_{\sigma}^r(t)}{\Delta}
 \left [ 
\begin{matrix} 
\Gamma_R & \Gamma_L\cr 
\Gamma_R & \Gamma_L
 \rule{0cm}{0.5cm}  \cr  
\end{matrix}          
 \right ] 
 +  
 \frac{G_{\sigma}^a(t)}{\Delta}
 \left [  
\begin{matrix} 
\ \Gamma_L & -\Gamma_L\cr 
-\Gamma_R & \ \Gamma_R
 \rule{0cm}{0.5cm}  \cr  
\end{matrix}          
 \right ] .
\label{eq:G_Vinf_intermediate_form_in_time}
\end{align} 
This is also one of the properties 
emerging in the high bias limit, 
where the density matrix is time independent.


\section{Charge and current fluctuations}

\label{sec:susceptibility}

One of the characteristics emerging in the 
high bias limit is a property of 
the commutation relation between 
the unperturbed and perturbed parts 
of the effective Hamiltonian
\begin{align} 
\left[\,
\widehat{H}_\mathrm{eff}^{(0)}
\, , \,
\widehat{H}_\mathrm{eff}^{(U)}
\,\right]
\,=\, i U \left( \,
 p_{\uparrow}^{} Q_{\downarrow}^{} 
+  p_{\downarrow}^{} Q_{\uparrow}^{} 
\right) \;.
\label{eq:commutation_H0_HU_Vinf}
\end{align}
Specifically, 
 $\widehat{H}_\mathrm{eff}^{(U)}$ and 
 $\widehat{H}_\mathrm{eff}^{(0)}$ commute 
in  a spin singlet subspace of the total charge $Q=0$, 
to which both the finial $\langle\!\langle I|\!|$ 
and the initial $|\! | \rho \rangle \! \rangle$ states belong. 
This affects significantly  the dynamics of the charges and currents  
which are determined by  
 the system of equations of motion given in 
  Eqs.\ \eqref{eq:EOM_Qtot_Vinf} and \eqref{eq:EOM_Qtot_Vinf_2}.
The $U$ dependent terms vanish from the equations of motion 
since the coefficients for the last equation of the four 
take the noninteracting values  
 $\mathcal{L}_{\sigma}^{2} = -\Delta^2$ 
and  $\mathcal{W}_{\sigma}^{} = \Gamma_L-\Gamma_R$ 
in the spin singlet subspace with $Q=0$.  
This is the central reason for the simplifications 
emerging in the high bias limit.

In order to see this in more detail, 
we consider the correlation function
\begin{align}
X_{AB}^{}(t,t')  
\,\equiv & \  
-i\, 
\langle\!\langle I|\!| \,\mathrm{T}\, 
A(t) \,
B(t') 
|\!|\rho(0) \rangle \! \rangle  \;,
\label{eq:correlation_function}
\end{align}
for the operators $A$ and $B$ 
which commute with the total charge $Q$, 
\begin{align}
\bigl[A\,,\, Q \bigr]\, = \, 
\bigl[B\,,\, Q \bigr] \, =\,0 \;.
\label{eq:Q_conservation_A_B}
\end{align}
Furthermore, we assume that both $A$ and $B$ satisfy the conditions:
\begin{align}
&
\langle\!\langle I|\!|
\bigl[
\mathcal{O}\,,\, 
\widehat{H}_{\mathrm{eff}}^{(U)}
\bigr]\,=\,0 \;, \qquad 
\bigl[
\mathcal{O}\,,\, 
\widehat{H}_{\mathrm{eff}}^{(U)}
\bigr]
|\! | \rho \rangle \! \rangle \,=\,0 \;,
\label{eq:charge_current_Q_conservation}
\end{align}
for $\mathcal{O}=A$ and $B$. 
 One of the most notable properties emerging in the high bias limit 
 is that $X_{AB}^{}(t,t')$  is not dependent on  $U$. 
This is because  the intermediate state for  $X_{AB}^{}(t,t')$ 
between the times $t$ and $t'$ also belong to the spin-singlet subspace 
with $Q=0$ owing to the first condition \eqref{eq:Q_conservation_A_B}. 
Furthermore, in this subspace, it is deduced 
from Eqs.\ \eqref{eq:commutation_H0_HU_Vinf} 
and \eqref{eq:charge_current_Q_conservation}  
that  $\widehat{H}_{\mathrm{eff}}^{(U)}$ 
commutes with $A$ and $B$ as well as $\widehat{H}_{\mathrm{eff}}^{(0)}$.
Therefore, the correlation function can be rewritten as   
\begin{align}
& 
X_{AB}^{}(t,0) 
\nonumber \\  
& \ \  = \,  
-i\langle\!\langle I|\!| \,\mathrm{T}\,  
e^{i\widehat{H}_{\mathrm{eff}}^{(0)} t}
e^{i\widehat{H}_{\mathrm{eff}}^{(U)} t}
A
e^{-i\widehat{H}_{\mathrm{eff}}^{(U)} t}
e^{-i\widehat{H}_{\mathrm{eff}}^{(0)} t}
\,
B
|\!|\rho (0)\rangle \! \rangle 
\nonumber \\
& 
\ \  =  \, 
-i \langle\!\langle I|\!| \,\mathrm{T}\,  
e^{i\widehat{H}_{\mathrm{eff}}^{(0)} t}
A
e^{-i\widehat{H}_{\mathrm{eff}}^{(0)} t}
\,
B
|\!|\rho \rangle \! \rangle 
\;.
\label{eq:susceptibilities_high_bias}
\end{align}
Here,  we set $t'=0$  for simplicity.
The conditions \eqref{eq:Q_conservation_A_B} and 
\eqref{eq:charge_current_Q_conservation} are satisfied 
for  the charges, currents and also spin operators. 
Thus, in the limit of $eV \to \infty$,   
the corresponding susceptibilities take 
their non-interacting forms, 
which can also be calculated from a particle-hole bubble 
in the Keldysh diagrammatic formalism.

\subsection{Charge and spin susceptibilities}

As an example, we consider
 the susceptibility for the impurity occupation 
$\delta n_{\mu,\sigma} \,\equiv\, n_{\mu,\sigma} 
- \langle n_{\mu,\sigma} \rangle$,
which satisfies those conditions given in  
Eq.\ \eqref{eq:charge_current_Q_conservation},
\begin{align}
\chi^{\mu\nu}_{\sigma\sigma'}(t) 
\,\equiv & \  
-i\, 
\mathrm{sign} (\mu) \,
\mathrm{sign} (\nu)\,
\langle\!\langle I|\!| \,\mathrm{T}\, 
\delta n_{\mu,\sigma}(t) \,\delta  n_{\nu,\sigma'} 
|\!|\rho \rangle \! \rangle  
\nonumber \\
 =& \
-i \langle\!\langle I|\!| \,\mathrm{T}\,  
\delta q_{\sigma}^{}(t) \,
\delta q_{\sigma'}^{} (0)
|\!|\rho \rangle \! \rangle 
\;.
\label{eq:correlation_function_physical_charge}
\end{align}
By definition, 
each  ($\mu$, $\nu$) component of this function 
is identical to the corresponding Keldysh correlation 
function with the same label.
The factor of $\mathrm{sign} (\mu)$ emerges through 
a similar way that $\bm{\tau}_3^{}$ 
appears in Eq.\ \eqref{eq:GKldysh_Gtfd}.
The second line of Eq.\  \eqref{eq:correlation_function_physical_charge} 
can be obtained by using an identity   
 $\delta n_{\mu,\sigma} 
 =
 \frac{1}{2}\,Q_{\sigma} - \mathrm{sign}(\mu) \,\delta q_{\sigma}^{}$ 
with $\delta q_{\sigma}^{} \equiv q_{\sigma}^{} 
- \langle q_{\sigma}^{} \rangle$. 
As in the case of Eq.\ \eqref{eq:susceptibilities_high_bias},
$\chi^{\mu\nu}_{\sigma\sigma'}(t)$ 
becomes $U$ independent for $eV \to \infty$   
and takes the noninteracting form,   
the Fourier transform of which is given by
\begin{align}
\chi^{\mu\nu}_{\sigma\sigma'}(\omega) 
 = & \ 
\frac{\Gamma_L \Gamma_R}{(\Gamma_L+\Gamma_R)^2}
\left(
\frac{1}{\omega +i2\Delta} - 
\frac{1}{\omega -i2\Delta}
\right) \, \delta_{\sigma\sigma'}
\;.  
\label{eq:qq_corelation}
\end{align}
Here,  the value of the imaginary part $2\Delta$  
in the denominator is determined by the operator  
 $\mathcal{L}_{\sigma}^{2}$ which takes 
eigenvalue $-\Delta^2$ in this case. 
As mentioned in Sec.\ \ref{subseec:current_algebra}, 
this eigenvalue causes a $e^{\pm 2\Delta t}$ dependence 
for the time-evolution of charges and currents 
through Eq.\ \eqref{eq:EOM_IL_Vinf}. 
Alternatively, in the diagrammatic approach 
this value emerges through the bubble diagram for 
the particle-hole pair excitation. 
It gives the total damping rate of $2\Delta$ 
in the limit of $eV \to \infty$ as a simple sum of the contributions 
of the particle and the hole parts.

We note that the charge and the spin susceptibilities 
are given by   
$\chi^{\mu\nu}_{\uparrow\uparrow} \pm \chi^{\mu\nu}_{\uparrow\downarrow}$.
Furthermore,  
Eq.\ \eqref{eq:qq_corelation}
indicates that 
all the  ($\mu$, $\nu$) components take the same value  
in the high bias limit, 
and thus the retarded susceptibilities vanish 
$
\chi^{r}_{\sigma\sigma'} \equiv
\chi^{--}_{\sigma\sigma'} -
\chi^{-+}_{\sigma\sigma'} \ \to \, 0$ in the high bias limit.

\subsection{Shot noise}

Another important example is the current noise, 
the spectrum of which can be deduced from the current-current 
correlation functions (in units of $e^2/\hbar$), 
\begin{align}
S_{\alpha\alpha'}(\omega) =  
\int_{-\infty}^{\infty} \!\! dt\, e^{i \omega t} 
\bigl[ \langle \delta I_{\alpha}(t) \, \delta I_{\alpha'}(0) \rangle  + 
 \langle \delta I_{\alpha'}(0) \,\delta I_{\alpha}(t)\rangle 
\bigl],
\end{align}
where 
$\delta I_{\alpha}(t)  \equiv   
\sum_{\sigma}
\bigl[
 I_{\alpha,\sigma}(t) -  \langle  I_{\alpha,\sigma}(t) \rangle 
\bigr]$ denotes the current fluctuations.
The current operators commute with the total charge as 
$\bigl[
I_{\alpha,\sigma} 
, 
Q_{\sigma'}^{}
\bigr]= 0$. Furthermore, they satisfy the conditions given in  
Eq.\ \eqref{eq:charge_current_Q_conservation} because   
$\bigl[
I_{R,\sigma} 
,
\widehat{H}_{\mathrm{eff}}^{(U)}
\bigr]=
U Q_{\overline{\sigma}}^{} I_{R,\sigma}$ and 
$\bigl[
I_{L,\sigma} 
, 
\widehat{H}_{\mathrm{eff}}^{(U)}
\bigr]=
-U Q_{\overline{\sigma}}^{} I_{L,\sigma}$.
Therefore, the current-current correlation 
functions are also not affected by the 
Coulomb interaction in the high bias limit 
as the one in  Eq.\ \eqref{eq:susceptibilities_high_bias}.
Thus, $S_{\alpha\alpha'}(\omega)$ also takes the non-interacting form 
\begin{align}
S_{RL}(\omega)= &  \, 
\frac{16\Gamma_L \Gamma_R}{(\Gamma_L+\Gamma_R)^2}
 \ i \left( 
\frac{ \Gamma_R^2}{\omega +i2\Delta} - 
\frac{ \Gamma_L^2}{\omega -i2\Delta}
\right) ,
\\ S_{LR}(\omega)= & \,  
\frac{16\Gamma_L \Gamma_R}{(\Gamma_L+\Gamma_R)^2}
 \ i \left( 
\frac{ \Gamma_L^2}{\omega +i2\Delta} - 
\frac{ \Gamma_R^2}{\omega -i2\Delta}
\right) , \\
S_{LL}(\omega)= &  
\left(\frac{4\Gamma_L \Gamma_R}{\Gamma_L+\Gamma_R}\right)^2
 (-i)
\left( 
\frac{1}{\omega +i2\Delta} - 
\frac{1}{\omega -i2\Delta}
\right) ,
\end{align}
and $S_{RR}(\omega) = S_{LL}(\omega) $.

\begin{figure}[t]
 \leavevmode
\begin{minipage}{0.35\linewidth}
\includegraphics[width=1\linewidth]{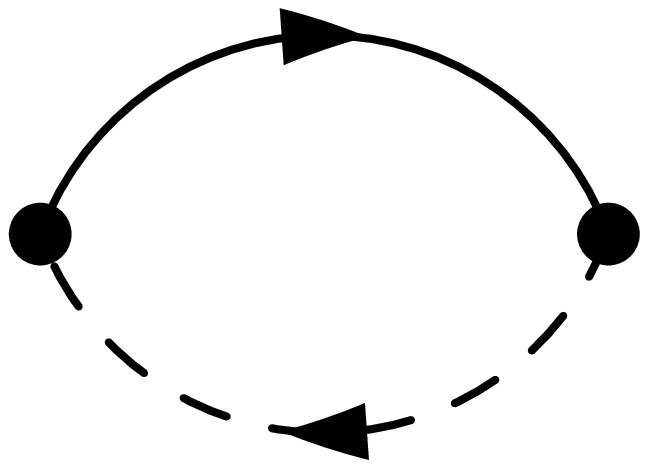}
\end{minipage}
\rule{0.08\linewidth}{0.0cm}
\begin{minipage}{0.35\linewidth}
\includegraphics[width=1\linewidth]{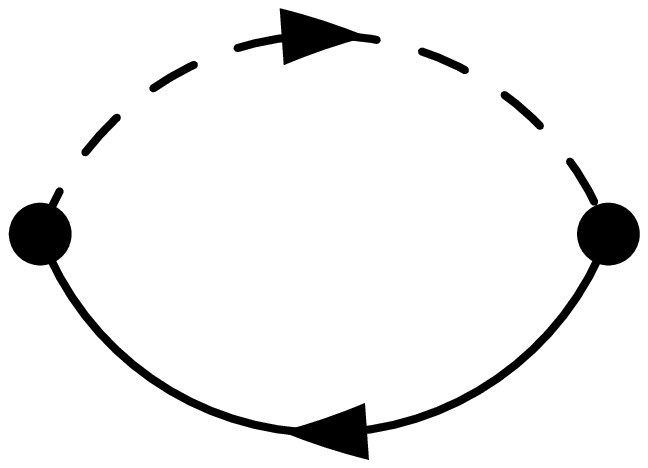}
\end{minipage}
\caption{
Feynman diagrams for $\delta S_{LL}^{}$ and  $\delta S_{RR}^{}$, 
which give additional contributions to the diagonal noise spectrum.  
The solid line denotes the dot Green's function 
$G^{\mu\nu}_{0,\sigma}$ 
and the dashed line denotes 
that of the isolated lead $g^{\mu\nu}_{\alpha}$ 
for $\alpha=L,\,R$.
}
 \label{fig:noise_add}
\end{figure}

There is one thing that we need to take into account for 
discussing the current-current correlation functions. 
In the Liouville-Fock space for the effective 
Hamiltonian $\widehat{H}_{\mathrm{eff}}^{}$,  
the operator equivalence between $I_{\alpha,\sigma}^{}$,   
defined in Eq.\ \eqref{eq:current_eV_inf_def}, 
and the corresponding original current vertex 
is sufficient for the off-diagonal components $S_{LR}$ and  $S_{RL}$. 
However, it is not faithful for obtaining 
the diagonal components corresponding to  $S_{LL}$ and  $S_{RR}$.
This is  because those processes which correspond to the bubble diagrams, 
consisting of   
 the Keldysh Green's function for the impurity site $\bm{G}_{0,\sigma}$ 
and that for the isolated lead  $\bm{g}_{\alpha}^{}$
as shown in Fig.\ \ref{fig:noise_add}, 
cannot be described by 
the operators, $I_{L,\sigma}^{}$ and $I_{R,\sigma}^{}$, projected into 
the finite Liouville-Fock space for $\widehat{H}_{\mathrm{eff}}^{}$. 
%
These processes can be systematically taken into account, 
 using the explicit expression of the current vertecies.
These additional contributions to the diagonal components,
for the current from the left lead, are given by  
\begin{align}
&\widetilde{S}_{LL}(\omega) 
\,= \,   
S_{LL}(\omega) + \delta S_{LL} \;, \\ 
&\delta S_{LL} 
\, = \,  
 \sum_{\sigma} 
\!\int_{-\infty}^\infty \! \frac{d\epsilon}{2\pi}
\,v_L^2 
\nonumber \\
& \qquad \times \Bigl[ \,
 g_{L}^{+-}(\epsilon)\,G_{0,\sigma}^{-+}(\epsilon+\omega)
+g_{L}^{-+}(\epsilon)\,G_{0,\sigma}^{+-}(\epsilon-\omega)
\nonumber \\
& \qquad \qquad 
+g_{L}^{-+}(\epsilon)\,G_{0,\sigma}^{+-}(\epsilon+\omega)
+g_{L}^{+-}(\epsilon)\,G_{0,\sigma}^{-+}(\epsilon-\omega)
\,\Bigr]
\nonumber \\
& \qquad 
\xrightarrow{eV\to\infty}  \ \ 
\frac{8\Gamma_L\Gamma_R}{\Gamma_L+\Gamma_R} \;. 
\end{align}
Here,
$g_L^{-+} = i 2\pi \rho_L f_L $, and 
$g_L^{+-} = -i 2\pi \rho_L (1-f_L)$. 
The distribution function takes 
the form $f_L(\epsilon) \to 1$ in the limit of $eV \to \infty$. 
Similarly, for the current to the right lead,
 we obtain  $\widetilde{S}_{RR}(\omega) 
=  S_{RR}(\omega) + \delta S_{RR}$ with $\delta S_{RR} \equiv \delta S_{LL}$. 
Therefore,  the zero-frequency shot noise is given by  
\begin{align}
S_{RL}(0) = & \ 
S_{LR}(0) = 
\widetilde{S}_{LL} (0) = 
\widetilde{S}_{RR} (0) 
\nonumber \\
= & \ 
\frac{4\Gamma_L^{} \Gamma_R^{} 
}{\Gamma_L^{} + \Gamma_R^{}}\, 
\left(1 + r^2
 \right) \ = \ \left( 1-r^4 \right) \Delta\;.
\end{align}
This also agrees with the noninteracting value 
that can be deduced from the Landauer-type formula 
in the limit of $eV\to\infty$ 
[see Appendix \ref{sec:some_expectation_values}].

\section{Exact Green's function for $U \neq 0$}
\label{sec:calculations_for_G}

In contrast to the correlation functions for the 
charges and currents discussed above, 
 the Green's function captures a non-trivial $U$ dependence. 
This is because the 
conditions described 
in Eqs.\ \eqref{eq:Q_conservation_A_B} and 
 \eqref{eq:charge_current_Q_conservation}
do not hold for the single-particle  and the single-hole excitations, 
and the  intermediate states belong 
to the subspace of $Q=1$ and $-1$. 
We provide in this section the exact interacting Green's function 
that is obtained in the high bias limit.

In the limit $eV \to \infty$,  
all the four components of the Green's function   
 can be expressed in terms of $G_{\sigma}^r$ and $G_{\sigma}^a$ 
as shown in  Eq.\ \eqref{eq:G_Vinf_intermediate_form_in_time}. 
Furthermore, the advanced function can be obtained from 
the retarded function through the Fourier transform  
 $G_{\sigma}^a(\omega) = \left\{G_{\sigma}^r(\omega)\right\}^*$.
Therefore, we concentrate on the retarded function, 
which can also expressed in the following form by  
using 
Eqs.\ \eqref{eq:bra_boundary} and \eqref{eq:Gr_Keldysh_sum_rule},
\begin{align}
G_{\sigma}^r (t) \, =&  \  
-i \,\theta(t)\,
\langle\!\langle I|\!| \,
d_{\mu, \sigma}^{}(t) \, 
\left(
d_{-,\sigma}^{\dagger} + 
d_{+,\sigma}^{\dagger}
\right)
|\! |  \rho  \rangle \! \rangle   
\;.
\label{eq:retarded_G_V_inf}
\end{align}
Here, the $G_{\sigma}^r$ in the left-hand side 
does not depend on whether the branch $\mu$ for the annihilation operator 
in the right-hand side is $-$ or $+$ 
owing to the properties of the final state 
$\langle\!\langle I|\!|$, shown in  Eq.\ \eqref{eq:bra_boundary}.

\subsection{Equations of motion for the retarded functions}

The Green's function can be exactly calculated  in the high bias limit 
from the Heisenberg representation given in Eq.\ \eqref{eq:G_eff_Vinf}, 
using the Lehmann representation 
in the finite Liouville-Fock space.\cite{SaptosovWegewijs}
 We provide, however, an alternative derivation using 
 the standard equation of motion approach that  can easily be  
compared with the well-known atomic limit solution.\cite{HubbardI}
The equation of motion for $G_{\sigma}^r$ can be derived 
simply by taking a derivative of 
Eq.\ \eqref{eq:retarded_G_V_inf} with  respect to $t$. 
 Then,  another correlation function appears through 
the commutation relation between $d_{\mu, \sigma}^{}$ and 
 $\widehat{H}_\mathrm{eff}^{(U)}$. 
Specifically, in the high bias limit, the dynamics of 
the charges and currents is determined by 
the closed system of equations, given in 
\eqref{eq:EOM_Qtot_Vinf} and \eqref{eq:EOM_Qtot_Vinf_2}.
We find that this also makes the equation of motion 
for $G_{\sigma}^r$ a closed system  
with two other composite correlation functions     
of $q_{\overline{\sigma}}^{}(t)$ and  $p_{\overline{\sigma}}^{}(t)$, 
 defined by 
%
\begin{align} 
&
\!\!\!
\Phi_{q;\sigma}^{r}(t) \equiv 
-i \,\theta(t) \langle\!\langle I|\!| \,
q_{\overline{\sigma}}^{}(t) \,
d_{\mu,\sigma}^{}(t)  \!
\left(
d_{-,\sigma}^{\dagger} \! + 
d_{+,\sigma}^{\dagger}
\right) \!
|\! |  \rho  \rangle \! \rangle, \!  
\\ 
 &
\!\!\!
\Phi_{p;\sigma}^{r}(t) \equiv  
-i \,\theta(t) \langle\!\langle I|\!| \,
p_{\overline{\sigma}}^{}(t) \,
d_{\mu,\sigma}^{}(t) \! 
\left(
d_{-,\sigma}^{\dagger}  \! + 
d_{+,\sigma}^{\dagger}
\right) \!
|\! |  \rho  \rangle \! \rangle  . \! 
\end{align}
Each of these two functions does not depend on 
whether $\mu = -$ or $+$ in the right-hand side, 
similarly to $G_{\sigma}^{r}$ shown in Eq.\ \eqref{eq:retarded_G_V_inf}. 
We obtain  such a 
closed system of equations of motion 
for $G_{\sigma}^{r}$, 
 $\Phi_{q;\sigma}^{r}$ and  $\Phi_{p;\sigma}^{r}$, 
after some straightforward calculations,  
\begin{widetext}
\begin{align}
&\left( i \frac{\partial}{\partial t} - \xi_{d,\sigma} 
+i\Delta 
\right)
G_{\sigma}^{r}(t)
-\, U 
\,\Phi_{q;\sigma}^{r}(t)
\, = \     \delta(t) 
\;, 
\label{eq:EOM_part1}
\\
& 
\left( i \frac{\partial}{\partial t} - \xi_{d,\sigma} 
+i\Delta 
\right)  
\Phi_{q;\sigma}^{r}(t)
+i \Phi_{p;\sigma}^{r}(t)
- \frac{U}{4} 
G_{\sigma}^{r}(t) 
\, =  \  
\frac{1}{2}\, 
 \frac{\Gamma_L-\Gamma_R}{\Gamma_L+\Gamma_R}
 \    \delta(t)  , 
\label{eq:EOM_part2}
\\ 
&\left( i \frac{\partial}{\partial t} - \xi_{d,\sigma} 
+ i\Delta \right)  
\Phi_{p;\sigma}^{r}(t)
\,-\, 2 \Delta \left[
\, \frac{U}{4} \,+\, 
i (\Gamma_L -\Gamma_R) \, 
\right]
  G_{\sigma}^{r}(t)
\,+\,i \Bigl[\,
4\,\Delta^2 
-i \, (\Gamma_L -\Gamma_R) U \, 
\Bigr] \,\Phi_{q;\sigma}^{r}(t)  
\, = \  0 
\;. 
\label{eq:EOM_part3}
\end{align}
\end{widetext}
This set of equations can be solved 
by carrying out the Fourier transformation,  
\begin{align}
 G_{\sigma}^{r} (\omega) 
\,=& \  
\frac{ \omega - \xi_{d,\sigma} +i3\Delta 
+ \,r\,\frac{U}{2}
}{\mathcal{D}_{\sigma}(\omega)}
\;, 
\label{eq:Vinf_retarded_in_w_new} 
\\ 
 \Phi_{q;\sigma}^{r}  (\omega) 
\,=& \  
\frac{1}{4}\,
\frac{ 
U+\,
2\,r\,  
(\omega - \xi_{d,\sigma} +i3\Delta) 
}{\mathcal{D}_{\sigma}(\omega)}
 \;, \\
 \Phi_{p;\sigma}^{r}  (\omega) 
\,=& \  
\frac{U}{2}\, 
\frac{  \left(1-r^2 \right)\Delta}{\mathcal{D}_{\sigma}(\omega)}
\;.
\end{align}
Here, 
$\mathcal{D}_{\sigma}(\omega)$ 
in the denominator is defined by 
\begin{align}
\mathcal{D}_{\sigma}(\omega) 
=  
 \left( \omega - \xi_{d,\sigma} +i3\Delta \right)
 \left( \omega - \xi_{d,\sigma} +i\Delta \right)
  - 
 \frac{U^2}{4} -i  r \Delta U.  
\end{align}
Here, $r$ is the parameter that 
describes the asymmetry in the couplings as 
shown in Eq.\ \eqref{eq:density_matrix_Vinf}, and 
 $\Delta = \Gamma_L+\Gamma_R$, as mentioned.

\subsection{Properties of the exact $G_{\sigma}^{r}(\omega)$}


We can also deduce the self-energy $\Sigma_{\sigma}^{r}(\omega)$ 
from  Eq.\ \eqref{eq:Vinf_retarded_in_w_new},  
which  can be  rewritten in  a similar 
form to the atomic-limit solution, 
\begin{align}
   G_{\sigma}^{r}(\omega)
  \, =&  \  \frac{1}{ \omega-\epsilon_{d,\sigma} 
    - \langle n_{d-\sigma} \,\rangle \,U  + 
    i\Delta  \,- \,  \Sigma_{\sigma}^{r}(\omega)}\;,
\label{eq:green_Vinf_result}
  \\
\nonumber\\
\Sigma_{\sigma}^{r}(\omega)
\,= & \  
\frac{
\langle n_{d-\sigma} \rangle\,
 \bigl(1-\langle n_{-\sigma}\rangle\bigr)\, U^2 
}{
\omega -\epsilon_{d,\sigma} \,-\,\bigl(1-\langle n_{d-\sigma}\rangle\bigr) 
\,U+  i3\Delta} \;, 
\label{eq:self_energy_Vinf_result}
\end{align}
where 
\begin{align}
 \langle n_{d-\sigma} \rangle \,\equiv\, \frac{\Gamma_L}{\Gamma_L+\Gamma_R} 
\ = \,  \frac{1+r}{2} \;. 
\end{align}
The asymmetry in the couplings $r$  varies 
the impurity occupation from half-filling.
We also see that  in the high bias limit 
 the self-energy $\Sigma_{\sigma}^{r}(\omega)$  
 has an imaginary part, $3\Delta$, in the denominator.
The original atomic-limit solution 
does not have such an imaginary part.\cite{HubbardI,Donianch}
This value, $3\Delta$, corresponds to 
a sum of the damping rates in the intermediate states, 
namely  $2\Delta$ of virtually excited one particle-hole pair,  
the propagation of which can be described by Eq.\ \eqref{eq:qq_corelation},  
and an additional $\Delta$  
of the incident particle (or hole).
 This value also coincides with the damping rate that  
one could phenomenologically expect from a Mathiessenn's rule 
for systems with a number of independent relaxation processes.    
In the special case of the symmetric couplings $r=0$, 
 the average impurity occupation is 
given by $ \langle n_{d-\sigma} \rangle =1/2$,  
and then the exact self-energy takes the same form 
as that of the order $U^2$ results,\cite{AO2002} namely 
 $\Sigma_{\sigma}^{r}(\omega) 
=  \left(\frac{U}{2}\right)^2/(\omega-\xi_{d,\sigma}+i3\Delta)$.
This is also exact for the high-temperature limit of thermal equilibrium, 
as mentioned.

\begin{figure}[t]
 \leavevmode
\begin{minipage}{0.9\linewidth}
\includegraphics[width=1\linewidth]{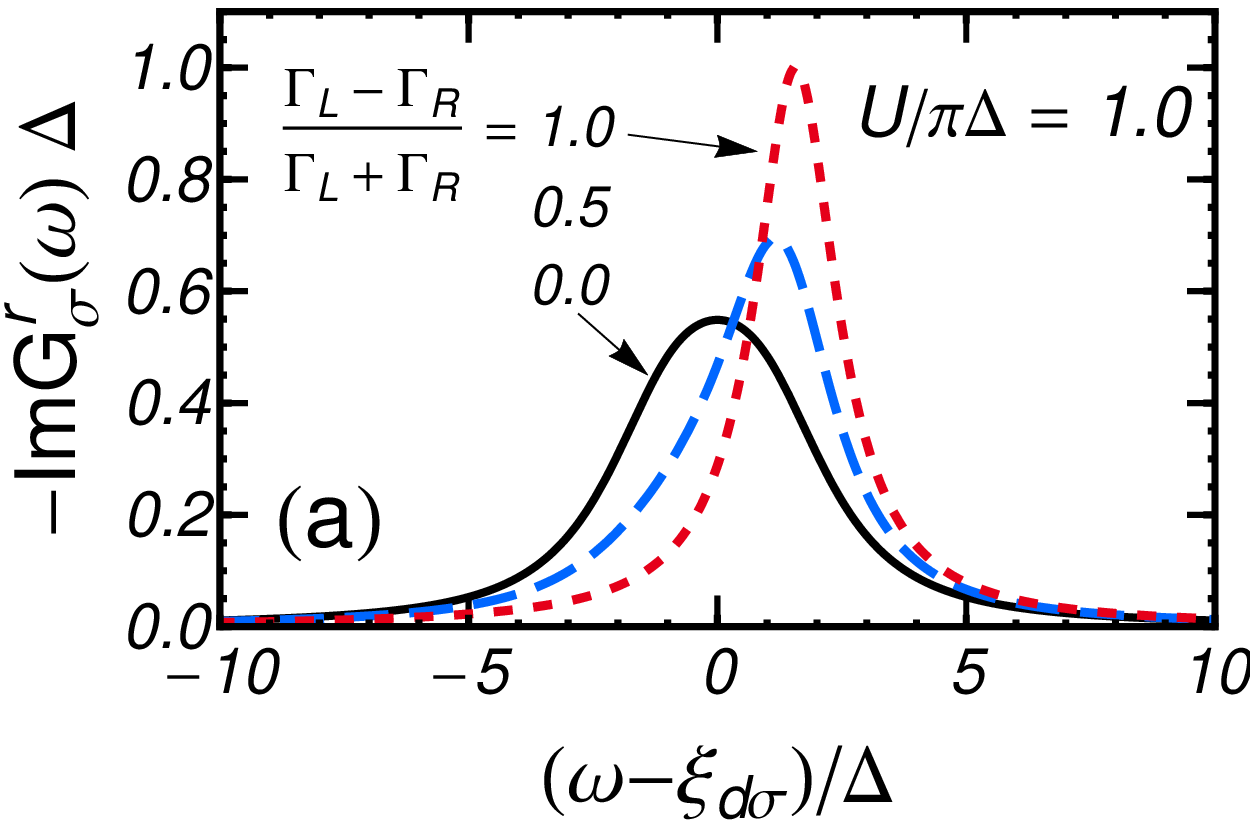}
\end{minipage}
\begin{minipage}{0.9\linewidth}
\includegraphics[width=1\linewidth]{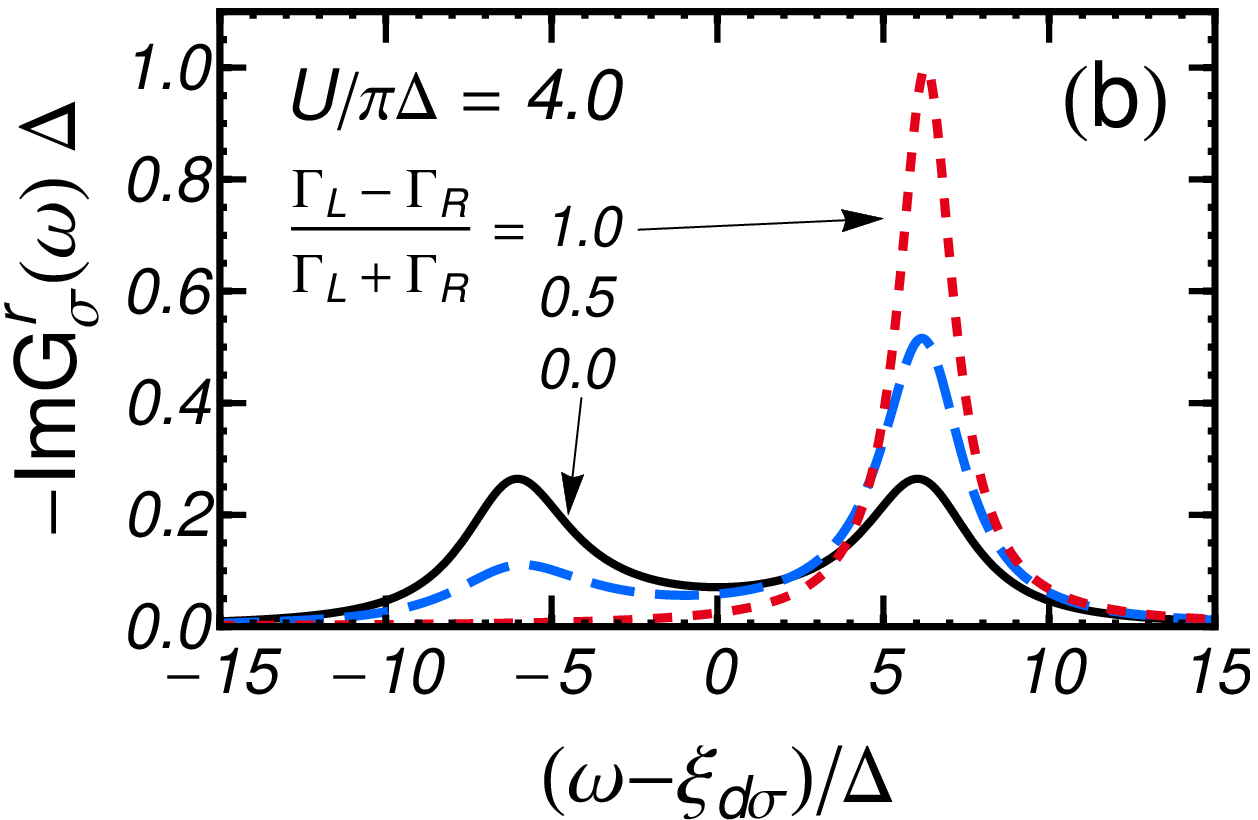}
\end{minipage}
\caption{(Color online) 
Spectral function at $eV \to \infty$ 
for $U/(\pi \Delta) =1.0$ (a) and  $4.0$ (b) 
for several values of the asymmetric couplings 
 $r \equiv (\Gamma_L-\Gamma_R)/(\Gamma_L+\Gamma_R)$.}  
 \label{fig:A_Vinf}
\end{figure}

Figure \ref{fig:A_Vinf} shows the spectral function, 
$-\mathrm{Im}\, G_{\sigma}^{r}(\omega)$, 
for several cases of the asymmetric couplings, 
choosing $U/(\pi \Delta)=1.0$ and $4.0$.
The spectral weight for the  upper (lower) Hubbard level 
increases for $r>0$ ($r<0$)   
as the asymmetry  in the couplings $|r|$ increases.
This peak approaches a Lorentzian form with the width $\Delta$ 
in the limit of $|r| \to 1$. 
Furthermore, in this figure, 
the separation of the upper and lower levels is clearly 
seen for strong interaction $U/(\pi \Delta)= 4.0$ 
while the peak structure is smeared 
for weak interaction $U/(\pi \Delta)= 1.0$. 
 Similarly,  for $r<0$,
the spectral weight of the lower Hubbard level increases.

%
%
%
%


The retarded Green's function,  given in 
Eq.\ \eqref{eq:Vinf_retarded_in_w_new}, 
can also be decomposed into the partial fractions,    
\begin{align}
&G_{\sigma}^r(\omega) 
 \,=  \   
\frac{Z_{+}^{}}{\omega-\xi_{d,\sigma}+i2\Delta-\mathcal{E}} 
+ 
\frac{Z_{-}^{}}{\omega-\xi_{d,\sigma}+i2\Delta+\mathcal{E}} 
\;,
\label{eq:Vinf_retarded_in_w}
\\
 &Z_{\pm}^{}  =      
\frac{1}{2} \left[ 1 \pm \frac{i\Delta +  
r \frac{U}{2}}{\mathcal{E}}
\right], \quad \, 
\mathcal{E} \equiv \sqrt{\left(\frac{U}{2}\right)^2 \!\! 
-  \Delta^2  + i r \Delta  U } . 
\end{align}
%
%
Here, the complex energy scale $\mathcal{E}$ determines 
the positions of the poles and 
its square $\mathcal{E}^2$ corresponds to  
the eigenvalue of the operator $\mathcal{L}_{\overline{\sigma}}^{2}$, 
 defined in Eq.\  \eqref{eq:L2_effective_def},  
for  $Q_{\sigma}^{} = 1$. 
 This energy scale emerges also as 
an eigenvalue of the effective Hamiltonian.\cite{SaptosovWegewijs}
The complex residues,  $Z_{+}^{}$ and $Z_{-}^{}$,  
represent the contributions of the upper and lower Hubbard levels,
which cause two different components emerging 
in the long-time behavior 
\begin{align}
&
\!\!\!\! 
G_{\sigma}^r(t) 
=    
-i\theta(t) 
\, e^{-i(\xi_{d,\sigma}-i2\Delta ) t}  
\Bigl(
 Z_+^{}\, e^{-i\mathcal{E}t}    
+Z_-^{}\,  e^{i\mathcal{E}t} 
\Bigr) .
\label{eq:Vinf_retarded_in_time_new}
\end{align}
Therefore, the relaxation time is determined by 
the imaginary part of 
the complex energy scale $\mathcal{E}$.\cite{SaptosovWegewijs}

\begin{figure}[t]
 \leavevmode
\begin{minipage}{0.9\linewidth}
\includegraphics[width=1\linewidth]{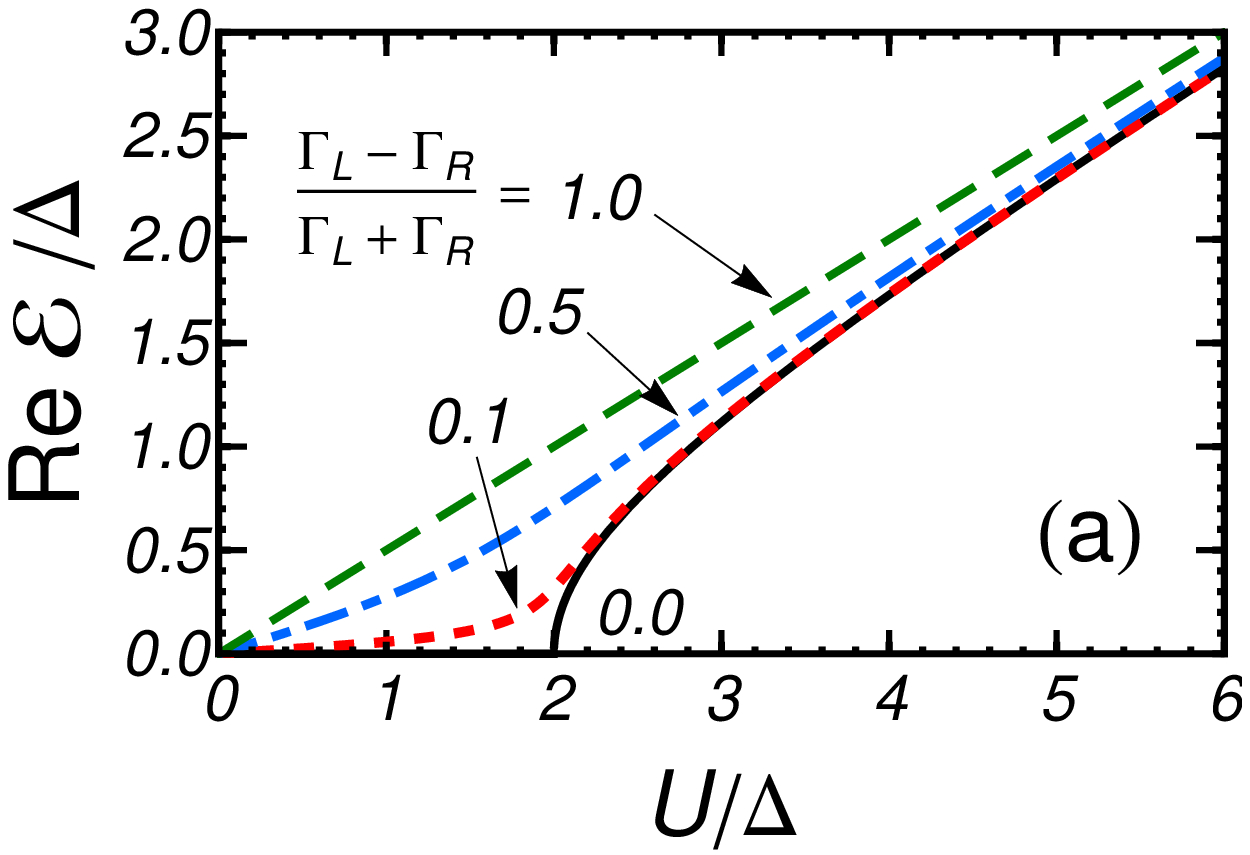}
\end{minipage}
\begin{minipage}{0.9\linewidth}
\includegraphics[width=1\linewidth]{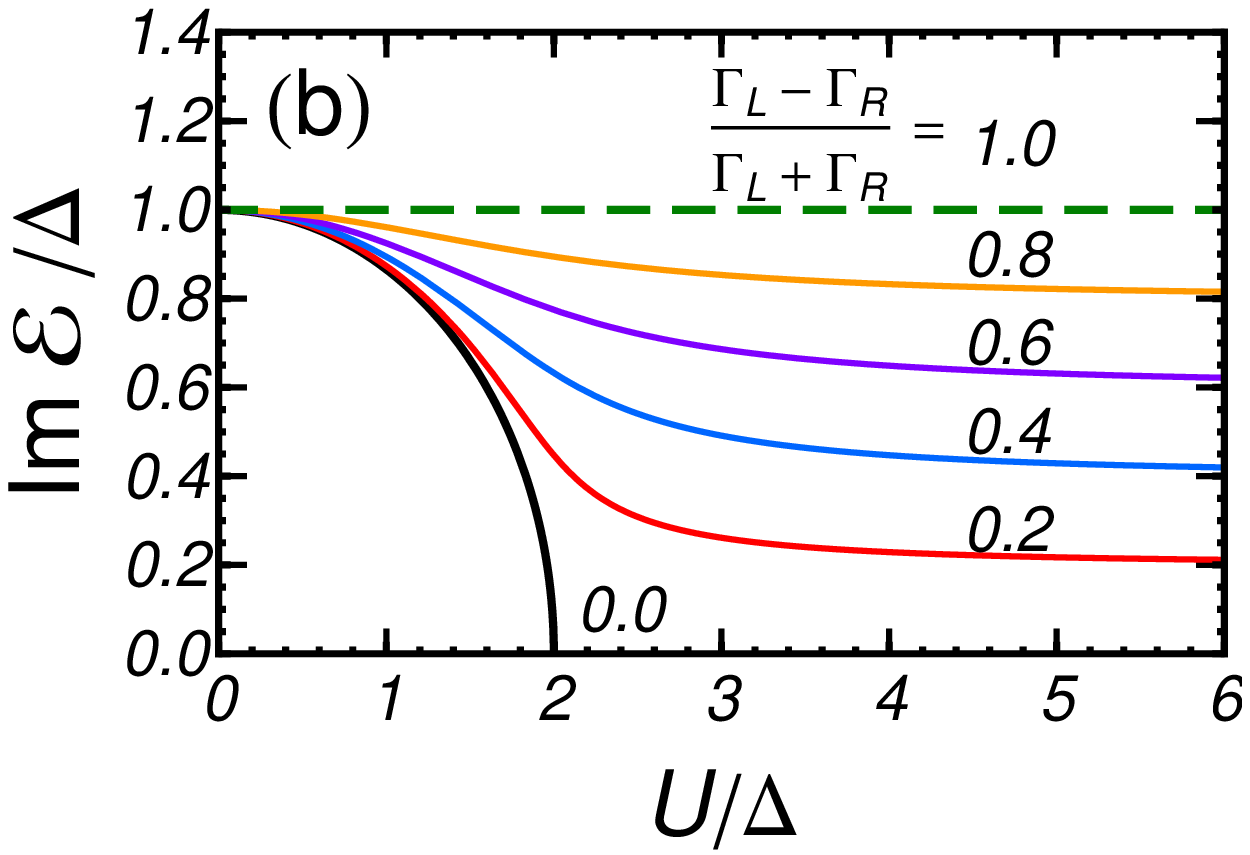}
\end{minipage}
\caption{(Color online) 
 $\mathrm{Re}\, \mathcal{E}$ and  $\mathrm{Im}\, \mathcal{E}$  
of the complex energy scale 
$\mathcal{E}  = 
\left[ 
\left(U/2\right)^2 -  \Delta^2 + i \,r\Delta  U
\right]^{1/2}$ 
are plotted vs $U$  
 for several values of $r \equiv (\Gamma_L-\Gamma_R)/(\Gamma_L+\Gamma_R)$ 
in (a) and (b), respectively.  
%
}
 \label{fig:imE}
\end{figure}

Figure\ \ref{fig:imE} shows the real and imaginary parts 
of $\mathcal{E}$ as a function of $U$ 
 for several cases of the asymmetric couplings $r$. 
The imaginary part of the complex energy is bounded 
in the range $|\mathrm{Im}\, \mathcal{E}| \leq \Delta$, and thus 
the relaxation rate of $G_{\sigma}^r(t)$ at long time 
takes the value of $2\Delta -  |\mathrm{Im}\, \mathcal{E}| \geq \Delta$. 
This indicates that  the relaxation becomes fast for $U > 0$  
whereas 
the asymmetry $|r|$ in the coupling makes relaxation slow.
 Note that in the noninteracting limit $U \to 0$, 
the complex energy scale and the 
residues approach $\mathcal{E} \to i \Delta$, 
 $Z_{+}^{} \to 1$, and   $Z_{-}^{} \to 0$.
For the symmetric couplings  $r=0$, 
the complex energy scale $\mathcal{E}$ becomes 
pure imaginary for $U<2\Delta$,  
and then it becomes real for $U>2\Delta$.
The asymmetry in the couplings, $r\neq 0$, makes this sudden change 
continuous, and  $\mathcal{E}$ takes the asymptotic 
form of  $\mathcal{E} \to U/2  +i \,r \Delta$ 
 for strong Coulomb repulsion  $U \gg \Delta$ 
while $\mathcal{E} \to |r|\, U/2  +i\, \mathrm{sign}(r)\, \Delta$ 
for weak interactions $U \ll \Delta$.
Furthermore, 
in the limit of $r = \pm 1$, where one of the leads is disconnected, 
the complex energy scale and 
the residues takes the values $\mathcal{E} = U/2  \pm i \Delta$, 
$Z_{\pm}^{}=1$, and $Z_{\mp}^{}=0$, respectively. 
%

\section{Summary}
\label{sec:summary}

We have presented the asymptotically exact solutions  for 
the interacting Green's function and the dynamic correlation 
functions for the charge, spin and current noise of the Anderson 
impurity in the high bias limit $eV \to \infty$ 
of a non-equilibrium steady state.

In this limit, the effective action 
along the Keldysh contour takes a simplified form  
 because the long-time tail of the kinetic part vanishes, 
as shown in Eq.\ \eqref{eq:S0_Vinf}. 
This form of the effective action can 
be replicated by the effective non-Hermitian Hamiltonian 
$\widehat{H}_\mathrm{eff}^{}$, 
given in Eqs.\ \eqref{eq:H_0_Vinf} and \eqref{eq:H_U_Vinf}, 
with the boundary conditions  described in  
Eqs.\ \eqref{eq:bra_boundary}--\eqref{eq:ket_boundary_2}.
The obtained effective Hamiltonian consists of the two sites, 
which correspond to the original impurity 
site and its image, defined with respect to 
the Liouville-Fock space in the context of 
the thermal field theory. 

It is found that $\widehat{H}_\mathrm{eff}^{}$ 
can  be expressed in terms of the 
charges and currents as Eq.\ \eqref{eq:H_Vinf_with_currents}.
This represents the essential characteristics of the high bias limit,  
and the equations of motion for the charges and the currents 
constitute a closed system, given in Eqs.\ 
\eqref{eq:EOM_Qtot_Vinf} and \eqref{eq:EOM_Qtot_Vinf_2}.
From this, it is deduced that 
the dynamic correlation functions for the operators 
which conserve the total charge    
become independent of $U$ in the high bias limit    
as shown in Sec.\ \ref{sec:susceptibility}.

It is also deduced from the properties of the effective Hamiltonian that 
all the components of the Keldysh Green's function 
are determined by a single element, for instance, 
by the retarded or advanced Green's function, 
as shown in Eq.\ \eqref{eq:G_Vinf_intermediate_form_in_time}.
Furthermore, owing to  the characteristics of the 
charge and current dynamics in the high bias limit, 
the equations of motion for the retarded Green's function 
$G_{\sigma}^{r}$ and the other two high-order correlation 
functions,  $\Phi_{q;\sigma}^{r}$ and  $\Phi_{p;\sigma}^{r}$,   
 constitute a closed system, 
given in Eqs.\ \eqref{eq:EOM_part1}--\eqref{eq:EOM_part3}.
The analytic solution for the Green's function 
and the self-energy can be expressed 
in the forms of  Eqs.\ \eqref{eq:green_Vinf_result} 
and \eqref{eq:self_energy_Vinf_result}, 
which can be compared with 
the atomic-limit solution\cite{HubbardI,Donianch,HaugJauho} 
and the order $U^2$ results.\cite{AO2002}
The results show that 
the self-energy generally captures a non-trivial 
imaginary part of the value of $3\Delta$ in the denominator 
in the high bias limit. 
Furthermore, the asymmetry in the couplings $\Gamma_L \neq \Gamma_R$  
varies the average impurity occupation and  
affects significantly the correlation effects due to $U$.

Since the seminal work of Hubbard,\cite{HubbardI}
the idea of the atomic limit has been applied and improved 
 in various fields of condensed matter physics.
For instance, it has recently been extended to study 
for quantum dots connected to superconductors.
\cite{Affleck,Vecino,sc2,TanakaNQDS,YsuhiroYamada}
This work  provides a new variety in the atomic-limit approaches.   
The exact results obtained at the high bias bounds   
will also be useful to explore further the nonequilibrium properties 
of quantum impurities in a wide energy range.


\begin{acknowledgments}

This work is supported by JSPS Grant-in-Aid 
for Scientific Research C (No.\ 23540375, and No.\ 24540316) 
and also that for Young Scientists (B) (No.\ 25800174).

\end{acknowledgments}

\appendix

\section{Some exact results at $eV \to \infty$}
\label{sec:some_expectation_values}

Using the usual Keldysh formalism, 
we show that the impurity occupation and 
the steady current do not depend on $U$ in the high bias limit.
First of all, the impurity occupation can be calculated, 
for instance, in the following way,\cite{Dutt}
\begin{align}
\langle n_{d,\sigma} \rangle 
\, = & \   
\int_{-\infty}^{\infty} 
\frac{d \omega}{2 \pi i} \, 
G_{\sigma}^{-+}(\omega) 
\, = \,   
\int_{-\infty}^{\infty} 
d \omega \, f_\mathrm{eff}(\omega)\,
A_{\sigma}(\omega) 
\nonumber \\
 \xrightarrow{eV \to\infty} &  \ 
\int_{-\infty}^{\infty} 
d \omega \,
 \frac{\Gamma_L}{\Gamma_R + \Gamma_L} 
\,A_{\sigma}(\omega) 
\ =  \ 
 \frac{\Gamma_L}{\Gamma_R + \Gamma_L} \;.
\end{align}
Note that the spectral function   
$A_{\sigma}(\omega) \equiv \left(- \frac{1}{\pi}\right) 
\mathrm{Im}\, G_{\sigma}^{r}(\omega)$ 
satisfies the sum 
rule $\int_{-\infty}^{\infty}  d \omega \,A_{\sigma}(\omega) =1$. 
Owing to this sum rule,  
the impurity occupation $\langle n_{d,\sigma} \rangle$  
becomes independent of 
the Coulomb repulsion $U$ and the level position $\epsilon_{d,\sigma}$   
in the high bias limit 
 although $A_{\sigma}(\omega)$ 
varies as a function of $U$ and $\epsilon_{d,\sigma}$. 
%
Similarly, in the high bias limit the current $J_{\sigma}$ 
due to the electrons with  spin $\sigma$  can be calculated as\cite{MW}  
\begin{align}
&
\!\!\! 
J_{\sigma} = 
\frac{e}{\hbar} \int_{-\infty}^{\infty}
\!\! \frac{d\omega}{2\pi}\,
\bigl[\, f_L(\omega) - f_R(\omega)\,\bigr] 
\,\frac{4\Gamma_L\Gamma_R}{\Gamma_L+\Gamma_R}
\,\pi A_{\sigma}(\omega) 
\nonumber \\
&
\xrightarrow{eV\to\infty}  
\frac{e}{\hbar} \! \int_{-\infty}^{\infty}\!\! d\omega\,
\, \frac{2\Gamma_L\Gamma_R}{\Gamma_L+\Gamma_R}
\, A_{\sigma}(\omega) 
 = 
\frac{e}{\hbar}
\, \frac{2\Gamma_L\Gamma_R}{\Gamma_L+\Gamma_R}
.
\end{align}
Here, the factor $e/\hbar$,   which  was assumed  
to be $1$ as a unit in the text,  is shown  explicitly.
One can also deduce the high-temperature value of the 
 thermal noise from the linear conductance, 
using the fluctuation-dissipation theorem,  
\begin{align}
4T  \frac{dJ}{dV} 
\, &  = \,    
 \frac{e^2}{\hbar}
\sum_{\sigma} 
    \frac{4\Gamma_L\Gamma_R}{\Gamma_L+\Gamma_R} \! 
\int_{-\infty}^{\infty} \!\! \frac{d\omega}{2\pi} \,
4T \left(\!- \frac{\partial f}{\partial \omega} \right) 
 \pi A_{\sigma}(\omega)
\nonumber \\ 
& \, \xrightarrow{T\to\infty}  
\frac{e^2}{\hbar}
\, \frac{4\Gamma_L\Gamma_R}{\Gamma_L+\Gamma_R}
\;.
\end{align}

In the  noninteracting case $U=0$, 
 the zero-frequency current noise 
can be expressed in the Landauer-type form 
\begin{align}
& 
\!\!\!\!\!
S_0(0) \,= \, 
\frac{e^2}{\hbar}
 \sum_\sigma 
\int \! \frac{d \omega}{2\pi} 
\  2 \biggl\{ \, 
\mathcal{T}_{\sigma} 
\, \left(\,1- \mathcal{T}_{\sigma} 
\,\right)
\,\left[\, f_L\, -\, f_R \,\right]^2  
\nonumber \\ 
&  \qquad \qquad \quad  
   +  
\mathcal{T}_{\sigma} 
\,  
\bigl[\, f_L\,(1-f_L) + (1-f_R)\,f_R \,\bigr] 
\biggr\} , 
\\
&
\!\!\!\!\!
\mathcal{T}_{\sigma}(\omega) 
\, \equiv \,  
\frac{4\Gamma_L^{} \Gamma_R^{}}{
 \left(\omega -\epsilon_{d,\sigma} 
\right)^2 
+ \left(\Gamma_L^{}+\Gamma_R^{} \right)^2 
} \;. 
\end{align}
From this expression, the value of 
the shot noise in the high bias limit $eV \to \infty$ 
can be deduced in the form  
\begin{align}
S_0(0) 
 \xrightarrow{eV\to\infty} &   
\, \frac{e^2}{\hbar}
\sum_\sigma 
\int_{-\infty}^{\infty} \! \frac{d \omega}{2\pi} 
\, 2 
\, 
\mathcal{T}_{\sigma} \, \left(\,1- \mathcal{T}_{\sigma} \,\right)
\nonumber \\
= & \ 
\frac{e^2}{\hbar}\, 
\frac{4\Gamma_L^{} \Gamma_R^{} 
}{\Gamma_L^{} + \Gamma_R^{}}\, 
\left[\,1 + 
\left(\frac{\Gamma_L^{} - \Gamma_R^{}
}{\Gamma_L^{} + \Gamma_R^{}}\right)^2
\, \right]
. 
\end{align}

\begin{figure}[t]
 \leavevmode
\begin{minipage}{0.5\linewidth}
%
\includegraphics[trim = 50 0 50 0, width=1\linewidth]{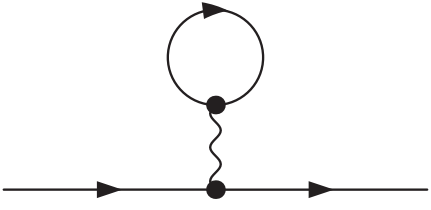}
\end{minipage}
\caption{
Feynman diagram for the Hartree term.}
 \label{fig:Hartree}
\end{figure}

\section{Feynman rule for the Hartree term}
\label{sec:Hartree}

There is a slight difference between  
the Feynman rules for the Keldysh Green's function 
 $G^{\mu\nu}_{\sigma}$ and 
  those for the Green's function  
 $\mathcal{G}^{\mu\nu}_{\sigma}$ defined with respect to 
the doubled Hilbert space. 
It emerges for the  $+$ component of 
the Hartree-type self-energy $\Sigma^{++}_{\sigma}$,
which corresponds to the tadpole diagram shown in Fig.\ \ref{fig:Hartree}.
As the arguments $t$ and $t'$ for the inner Green's function 
along the loop are equal, 
the limit is required to be taken carefully  
 such that $G^{++}_{\overline{\sigma}}(t+ 0^+,t)$ 
in the Keldysh approach  
 whereas the opposite limit is required 
for $\mathcal{G}^{++}_{\overline{\sigma}}(t,t+ 0^+)$ 
in the thermal-field-theoretical approach.  
  This is caused by the difference 
in the direction of the time-ordering for 
the operators belonging to the $+$ branch. 
Thus, for the $-$ component of
the Hartree-type self-energy $\Sigma^{--}_{\sigma}$,
the same limit $t'\to t+0^+$ is taken  
for both the Keldysh 
and the thermal-field-theoretical Green's functions.

The effective Hamiltonian $\widehat{H}_\mathrm{eff}^{}$, 
defined in Eqs.\ \eqref{eq:H_0_Vinf} and \eqref{eq:H_U_Vinf}, 
includes the $U$-dependent terms such that  
\begin{align}
\frac{U}{2} \sum_{\sigma} Q_\sigma^{} \! 
+ \widehat{H}_\mathrm{eff}^{(U)}  
 = &  \  
U \left( n_{-,\uparrow} \,n_{-,\downarrow}  -   
n_{+,\uparrow}\,n_{+\downarrow} \right)
+ \widehat{H}_\mathrm{eff}^{(\mathrm{CT})}  ,
\nonumber \\
 \widehat{H}_\mathrm{eff}^{(\mathrm{CT})}  
\equiv & \  
U \left( n_{+,\uparrow} + n_{+,\downarrow}  -1 \right) 
\;, 
\end{align}
Therefore, the last term, $\widehat{H}_\mathrm{eff}^{(\mathrm{CT})}$,  
also appears in the perturbation expansion 
with respect to $U$ in  the thermal-field-theoretical approach. 
This term can be regarded as a counter term 
for the particles in the $+$ branch,
and compensates the difference that 
 arises in the $+$ component of the Hartree energy shift, 
mentioned above. 
We also note that, by definition, the relation between 
the Keldysh self-energy 
$\bm{\Sigma}_{\sigma}^{}$
for $\bm{G}_{\sigma}$ and the self-energy 
$\bm{\mathit{\Sigma}}_{\sigma}^\mathrm{TFT}$
for $\bm{\mathcal{G}}_{\sigma}$ 
can be expressed in the form  
\begin{align}
\bm{\Sigma}_{\sigma}^{}(\omega)
\, = \, \bm{\tau}_3 \,\bm{\mathit{\Sigma}}_{\sigma}^\mathrm{TFT}(\omega)\;.
\end{align}


\begin{thebibliography}{99}

\bibitem{Anderson}
P.\ W.\ Anderson, Phys.\ Rev.\  {\bf 124}, 41 (1961).


\bibitem{HDW}
 S.\ Hershfield, J.\ H.\ Davies, and J.\ W.\ Wilkins, 
 Phys.\ Rev.\ B {\bf 46}, 7046 (1992).


\bibitem{MW} 
Y.\ Meir  and  N.\ S.\ Wingreen, 
Phys.\ Rev.\ Lett.\ {\bf 68}, 2512 (1992).


\bibitem{DMFT}
 A.\ Georges, G.\ Kotliar, W.\ Krauth, and M.\ J.\ Rozenberg,
 Rev.\ Mod.\ Phys.\ {\bf 68}, 13 (1996).


\bibitem{Hewson_book}
A.\ C.\ Hewson, 
 { \em  The Kondo Problem to Heavy Fermions\/} 
 (Cambridge University Press, Cambridge, 1993). 


\bibitem{KNG}
 A.\ Kaminski, Yu.\ V.\ Nazarov, and L.\ I.\ Glazman, 
 Phys.\ Rev.\ B {\bf 62}, 8154 (2000).


\bibitem{ao2001}
A.\ Oguri, Phys.\ Rev.\ B {\bf 64}, 153305 (2001).


\bibitem{HBA} 
 A.\ C.\ Hewson, J.\ Bauer, and A.\ Oguri,
 J.\ Phys.: Condes.\ Matter.\  {\bf 17}, 5413 (2005).




\bibitem{GogolinKomnik} 
A.\ O.\ Gogolin and A.\ Komnik, 
 Phys.\ Rev.\ B {\bf 73}, 195301 (2006).


\bibitem{Sela2006} 
E.\ Sela, Y.\ Oreg, F.\ von Oppen and J.\ Koch, 
Phys.\ Rev.\ Lett.\ {\bf 97}, 086601 (2006)


\bibitem{Golub} 
A.\ Golub, Phys.\ Rev.\ B {\bf 73}, 233310 (2006).

\bibitem{Mora2009}
C.\ Mora, P.\ Vitushinsky, X.\ Leyronas, A.\ A.\ Clerk, and 
K.\ Le Hur, Phys. Rev. B {\bf 80}, 155322 (2009).



\bibitem{Fujii2010}
T.\ Fujii, J.\ Phys.\ Soc.\ Jpn.\ {\bf 79}, 044714 (2010). 


\bibitem{Sakano}
 R.\ Sakano, T.\ Fujii, and A.\ Oguri, 
 Phys.\ Rev.\ B {\bf 83}, 075440  (2011).




\bibitem{Grobis}
M.\ Grobis, 
I.\ G.\ Rau, R.\ M.\ Potok, H.\ Shtrikman, 
and D.\ Goldhaber-Gordon,
Phys.\ Rev.\ Lett.\ {\bf 100}, 246601 (2008).


\bibitem{ScottNatelson}
G.\ D.\ Scott, 
 Z.\ K.\ Keane, J.\ W.\ Ciszek, J.\ M.\ Tour, and D.\ Natelson, 
Phys.\ Rev.\ B {\bf 79}, 165413 (2009).



\bibitem{Heiblum} 
O.\ Zarchin, M.\  Zaffalon,  M.\ Heiblum, D.\ Mahalu, and V.\ Umansky, 
Phys.\ Rev.\ B {\bf 77}, 241303  (2008).


\bibitem{Kobayashi}
Y.\ Yamauchi, K.\ Sekiguchi, K.\ Chida, T.\ Arakawa, S.\ Nakamura, 
K.\ Kobayashi, T.\ Ono, T.\ Fujii, and R.\ Sakano, 
Phys.\ Rev.\ Lett.\ {\bf 106}, 176601 (2011).




 \bibitem{Anders}
  F.\ B.\ Anders, Phys.\ Rev.\ Lett.\ {\bf 101}, 066804 (2008).

 \bibitem{Kirino}
  S.\ Kirino, T.\ Fujii, J.\ Zhao, 
 and K.\ Ueda, J.\ Phys.\ Soc.\ Jpn {\bf 77}, 084704 (2008).

 \bibitem{Werner}
  P.\ Werner, T.\ Oka, and A.\ J.\ Millis, Phys. Rev. B {\bf 79}, 
 035320 (2009).

 \bibitem{MuhlbacherUrbanKomnik}
L.\ M\"{u}hlbacher, D.\ F.\ Urban, and A.\ Komnik
Phys.\ Rev.\ B {\bf 83}, 075107  (2011). 

 \bibitem{Han}
  J.\ E.\ Han, A.\ Dirks, and T.\ Pruschke, Phys. Rev. B {\bf 86}, 
 155130 (2012).




 \bibitem{Keldysh} 
 L.\  V.\ Keldysh, 
 Sov.\ Phys.\ JETP {\bf 20},  1018 (1965) 
[Zh.\ Eksp.\ Teor.\ Fiz.\ {\bf 47}, 1515 (1964)].


 \bibitem{Caroli} 
 C.\ Caroli, R.\ Combescot, P.\ Nozieres, and D.\ 
 Saint-James, J.\  Phys.\ C {\bf 4},  916 (1971).


\bibitem{UmeMatTac}
 H.\ Umezawa, H.\ Matsumoto, and M.\ Tachiki, 
 {\it Themo Field Dynamics and Condensed States\/} 
 (North-Holland, Amsterdam, 1982).



\bibitem{EzaAriHas}
H.\ Ezawa, T.\ Arimitsu, and Y.\ Hashimoto, 
{\it Themal Field Theories\/} 
(North-Holland, Amsterdam, 1991).

 





\bibitem{HubbardI}
J.\ Hubbard, Proc.\ Roy.\ Soc.\ {\bf A276}, 238  (1963).

\bibitem{Donianch}
S.\ Doniach, Adv.\ Phys.\ {\bf 18}, 819  (1969).

\bibitem{HaugJauho}
H.\ Haug and A.\ -P.\ Jauho
  { \em  Quantum Kinetics in Transport and Optics of Semiconductors  \/} 
  (Springer, Berlin, 1996). 


\bibitem{AO2002}
A.\ Oguri, J.\ Phys.\ Soc.\ Jpn.\ {\bf 71}, 2969 (2002).







 


\bibitem{EspositeHarolaMukamel}
M.\ Esposite, U.\ Harbola, and S.\ Mukamel,
Rev.\ Mod.\ Phys.\ {\bf 81}, 1665  (2009).


\bibitem{DzioevKosov}
A.\ A.\ Dzhioev and D.\ S.\ Kosov,
J.\ Chem.\ Phys.\ {\bf 34}, 154107 (2011).


\bibitem{SaptosovWegewijs}
R.\ B.\ Saptsov and M.\ R.\ Wegewijs,
 Phys.\ Rev.\ B {\bf 86}, 235432 (2012).



\bibitem{GL}
Y.\ Dzyaloshinskii, and A.\ I.\ Larkin, 
Soviet Phys.\  JETP {\bf 38}, 202 (1974)
[Zh.\ Eksp.\ Teor.\ Fiz.\ {\bf 65}, 411 (1973)].

\bibitem{ES}
E.\ U.\ Everts and H.\ Shultz, Solid State Commun.\ 
{\bf 15}, 1413 (1974).




\bibitem{Affleck}
I.\ Affleck, J.\ -S.\ Caux, and A.\ M.\ Zagoskin,
Phys.\ Rev.\ B \textbf{62}, 1433 (2000).

\bibitem{Vecino}
E. Vecino, A. Martin-Rodero, and A. Levy Yeyati,
Phys.\ Rev.\ B \textbf{68}, 035105  (2003).

\bibitem{sc2} 
Yoshihide Tanaka,  A.\ Oguri, and A.\ C.\ Hewson, 
 New J.\ Phys.\ {\bf 9},  115 (2007);  
 {\bf 10}, 029801(E) (2008).

\bibitem{TanakaNQDS}
Yoichi Tanaka, N.\ Kawakami, and A.\ Oguri,
J.\ Phys.\ Soc.\ Jpn.\ \textbf{76}, 074701 (2007);
\textbf{77}, 098001(E) (2008).


\bibitem{YsuhiroYamada}
Yasuhiro Yamada, 
 Yoichi Tanaka, and N.\ Kawakami, 
Phys.\  Rev.\ B {\bf 84}, 075484 (2011).




\bibitem{Dutt}
P.\ Dutt, J.\ Koch, J.\ E.\ Han, Karyn Le Hur, 
Ann.\ Phys.\ {\bf 326}, 2963 (2011). 








\end{thebibliography}
\end{document}